\documentclass[onecolumn]{aastex63}
\usepackage{mathptmx}
\usepackage[T1]{fontenc}
\usepackage[letterpaper]{geometry}
\usepackage{xcolor}
\usepackage{graphicx}
\usepackage{amsmath}

\makeatletter
\usepackage{times}
\makeatother

\usepackage{babel}
\begin{document}

\title{Plumelets: Dynamic Filamentary Structures in Solar Coronal Plumes}

\author{V.~M. Uritsky}
\affil{Catholic University of America, 620 Michigan Avenue NE, Washington DC 20061, USA}
\affil{NASA Goddard Space Flight Center, 8800 Greenbelt Road, Greenbelt MD 20771, USA}

\author{C.~E. DeForest}
\affil{Southwest Research Institute, 1050 Walnut Street Suite 300,
Boulder CO 80302, USA}

\author{J.~T. Karpen}
\author{C.~R. DeVore}
\affil{NASA Goddard Space Flight Center, 8800 Greenbelt Road, Greenbelt MD 20771, USA}

\author{P. Kumar}
\affiliation{Department of Physics, American University, Washington, DC 20016, USA}
\affiliation{NASA Goddard Space Flight Center, 8800 Greenbelt Road, Greenbelt MD 20771, USA}

\author{N.~E. Raouafi}
\affil{Johns Hopkins University Applied Physics Laboratory, 11100 Johns Hopkins Road, Laurel MD 20723, USA}

\author{P.~F. Wyper}
\affil{Durham University, Durham DH1 3LE, UK}

\begin{abstract}

Solar coronal plumes long seemed to possess a simple geometry supporting spatially coherent, stable outflow without significant fine structure. Recent high-resolution observations have challenged this picture by revealing numerous transient, small-scale, collimated outflows (``jetlets'') at the base of plumes. The dynamic filamentary structure of solar plumes above these outflows, and its relationship with the overall plume structure, have remained largely unexplored. We analyzed the statistics of continuously observed fine structure inside a single representative bright plume within a mid-latitude coronal hole during 2016 July 2-3. By applying advanced edge-enhancement and spatiotemporal analysis techniques to extended series of high-resolution images from the \emph{Solar Dynamics Observatory}'s Atmospheric Imaging Assembly, we determined that the plume was composed of numerous time-evolving filamentary substructures, referred to as ``plumelets'' in this paper,  that accounted for most of the plume emission. The number of simultaneously identifiable plumelets was positively correlated with plume brightness, peaked in the fully formed plume, and remained saturated thereafter. The plumelets had transverse widths of 10 Mm and intermittently supported upwardly propagating periodic disturbances with phase speeds of 190-260 km s$^{-1}$ and longitudinal wavelengths of 55-65 Mm. The characteristic frequency (3.5 mHz) is commensurate with that of solar $p$-modes. Oscillations in neighboring plumelets are uncorrelated, indicating that the waves could be driven by $p$-mode flows at spatial scales smaller than the plumelet separation. Multiple independent sources of outflow within a single coronal plume should impart significant fine structure to the solar wind that may be detectable by \emph{Parker Solar Probe} and \emph{Solar Orbiter}.

\end{abstract}

\section{Introduction}\label{sec:intro}

Solar plumes have been observed since  humans first gazed in wonder at solar eclipses, yet their origin and dynamics are not yet thoroughly understood \citep[see reviews by][]{wilhelm2011,poletto2015}. These common features are bright collimated structures within darker, larger coronal holes; after formation, they persist from hours to days. Plumes are both denser and cooler than their surroundings, and emanate from nearly unipolar magnetic-flux concentrations at the photosphere. At sufficiently high spatial resolution and temporal cadence, plumes are not monolithic, stationary structures. Instead, they exhibit both transverse fine structure and dynamic features that travel along bright threads. The nature of the propagating disturbances remains controversial, as it is difficult to determine whether they represent bulk motions, waves, or some combination of both.

Early evidence for filamentary structures within plumes at 10$''$ in size was reported from analyses of data from \emph{Skylab} \citep{karovska1994} and the \emph{Solar and Heliospheric Observatory} \citep[\emph{SOHO;}][]{deforest1997,deforest1998}. \citet{ofman97} and \citet{deforest1998} measured oscillations in plumes with periods of several minutes and propagation speeds above 100 km s$^{-1}$, which they interpreted as evidence for compressive sound or slow magnetoacoustic waves. \citet{ofman1999} replicated key features of the observations with analytical and numerical calculations of magnetohydrodynamic waves in a model plume. Based on these and other observations, \citet{wang1998} proposed that plumes result from interchange magnetic reconnection between the ambient unipolar magnetic field of coronal holes and bipolar flux emerging from below the photosphere.

Subsequent observations from \emph{Hinode} and the \emph{Solar - Terrestrial Relations Observatory} (\emph{STEREO}) revealed abundant small-scale structures and dynamics at the base of plumes. \citet{raouafi2008} found strong associations between polar coronal-hole jets and/or coronal bright 
points and the formation/enhancement of  plumes. \citet{gabriel2009} concluded that diffuse, large-scale plumes are aggregates of numerous smaller-scale plumes. Applying wavelet analysis to coronal images of plumes from the \emph{Solar Dynamics Observatory} (\emph{SDO}) enabled \citet{raouafi2014} to extract even finer-scale structures and dynamics, which they referred to as ``jetlets'' and plume transient bright points. These features are reduced-scale versions of the jets and bright points that had been suggested previously as sources of plume plasma.  Observations show that coronal jets and plumes could be connected in at least two different ways. X-ray jets precede some plumes \citep{raouafi2008}, while small-scale jets (i.e., jetlets) may sustain long-term plume evolution \citep{raouafi2014}. This dynamic activity was associated with mixed magnetic polarity, although the large-scale magnetic flux at plume footpoints is dominantly unipolar. On the other hand, more recent studies have found that most coronal-hole jets do not result in plumes \citep[e.g.,][]{kumar2019}, so the relationship remains unclear.

Other recent work has reinforced the association between the plumes and small-scale footpoint activity, including investigations of a bright-point/plume association over a 40-hour interval \citep{pucci2014}, combined \emph{SDO} and \emph{Interface Region Imaging Spectrograph} (\emph{IRIS}) observations of tiny jets and bright points \citep{pant2015}, and strong correlations between plume emission and converging flows in the magnetic network \citep{wang2016}. In the last study, small-scale closed loops were observed at the base of plumes even when the minority-polarity magnetic flux was weak, diffuse, or undetectable, suggesting that mixed magnetic polarity is a crucial ingredient in the activation and maintenance of plumes.

To gain further insights intothe mass source and physical mechanism generating plumes, we have applied state-of-the-art image-processing techniques \citep{deforest_fading_2016,deforest_noise-gating_2017} to a long-duration, high-resolution, high-cadence sequence of \emph{SDO} observations of a plume in a low-latitude coronal hole on 2016 July 2-3. Our detailed, quantitative analysis of this unprecedented data set has yielded key new information on the plume filamentary structures, which we refer to as ``plumelets,'' and their quasiperiodic oscillations. 

We find that our coronal plume contains well-developed transverse structure, which is masked by diffuse background emission and becomes evident after proper image processing. The number of detectable plumelet edges increases systematically with the average plume brightness, reaching more than 30 once the plume is fully formed. The plumelets exhibit a persistent pattern of upward propagating quasiperiodic disturbances, which could be a manifestation of repetitive plasma outflows, slow mode waves, or both. The frequency of the disturbances is consistent with that of solar $p$-mode oscillations, while the phases of the disturbances observed in different plumelets are effectively random. Our analysis indicates that plumelets could be formed by bursty reconnection in a topologically complex lower corona, driven by a combination of random and periodic photospheric motions. 

Plumelets, as defined in this paper, represent a fundamental and persistent attribute of the coronal-hole plume that we observed, and may well be common to all plumes. For instance, a close look at the images of three plumes analyzed by \citet{raouafi2014} reveals ubiquitous plumelet-like structures similar those studied here. Although such fine structure has been known to exist in plumes for a long time, we are not aware of any previous studies focusing on their quantitative properties such as their characteristic spatial and temporal scales, or on their relationship with the long-term evolution of the entire plume in which they are embedded. Our present paper addresses these important aspects of coronal plumelets based on a rigorous statistical analysis of an extended set of high-resolution images of a representative plume system observed over several consecutive days.
We are currently analyzing the impulsive jets and bright points at all scales at the base of this plume to determine their connections to the plumelets \citep[][in preparation]{kumar2020}. 

We present the data set and its processing in \S2, describe our results in \S3, discuss their implications in \S4, and provide our conclusions in \S5.

\section{Data}\label{sec:data}

\begin{figure}[tbh]
\begin{centering}
\includegraphics[width=6.5in]{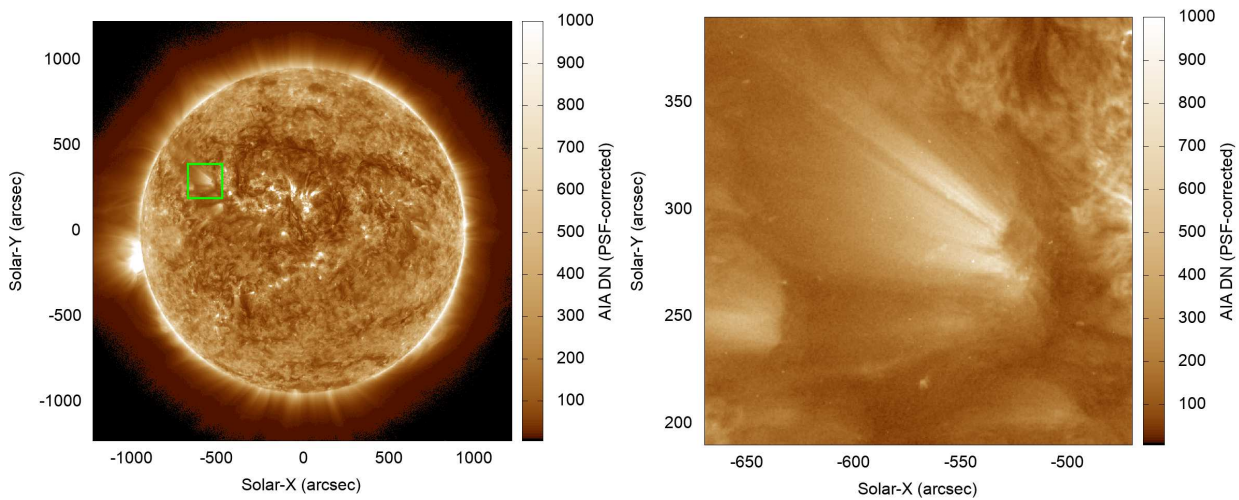}
\par\end{centering}
\caption{\label{fig:171-Context}(Left) \emph{SDO}/AIA 171\,\AA~images of the full Sun on 2016 July 2 at 19:35 UT, with the region of interest outlined by a green box. (Right) The region of interest containing the selected plume. We analyzed the plume evolution from 2016 July 2 00:00 UT through 2016 July 3 17:00 UT; this image is near the center of our study, in Interval 5 (below). }
\end{figure}

Figure \ref{fig:171-Context} reproduces two Level-1 (L1) images that show the overall corona, coronal hole, and location of the plume that we examined, which was observed from 2016 July 2 00:00 UT through 2016 July 3 17:00 UT. A large data gap followed this observing period, and by July 4 the plume was too diffuse to yield useful results. We used the full-time-resolution image sequence from \emph{SDO}/AIA \citep{lemen_atmospheric_2011,lemen_atmospheric_2012} in the 171\,\AA, 193\,\AA, and 211\,\AA~channels, post-processed to enhance sensitivity. Plumes are bright in these wavelengths, whose collisional-excitation temperature responses peak in the 0.8-2.0 MK range that dominates in most cool coronal structures (\citealt{deforest_multi-spectral_1991,odwyer_sdo/aia_2010,lemen_atmospheric_2011}). We also examined the 335\,\AA~channel (Figure \ref{fig:Close-up-images-of}, lower-right panel) but the count rates were negligible ($\lesssim 7$ counts per frame at the plume base), so we excluded those images from further analysis. The data were post-processed in four distinct steps:

\begin{enumerate}
\item We convolved each image frame with the corresponding inverse point-spread function (PSF) calculated for that channel \citep{poduval_point-spread_2013}, 
then cropped the frame to the region around the selected plume observed on 2016 
July 4. Convolving with the calculated inverse PSF is equivalent to the iterative 
deconvolution method described by \citet{grigis_aia_2011}, because images form a 
semigroup under convolution. The deconvolution step reduces ``haze'' by removing predetermined/calibrated stray light from the images, and requires the full frame 
before cropping. Figure \ref{fig:Close-up-images-of} shows the resulting images 
of the plume near the beginning of the data set on 2016 July 2.

\begin{figure}[tbh]
\begin{centering}
\includegraphics[width=6.25in]{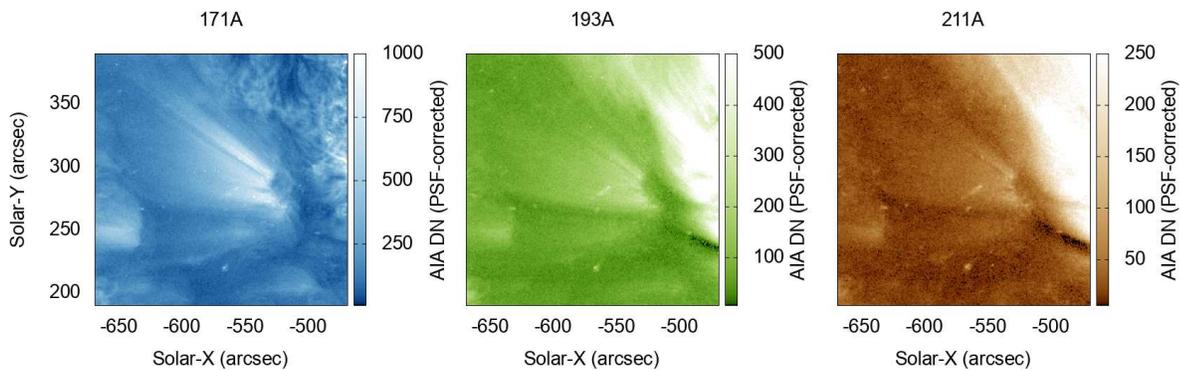}
\par\end{centering}
\caption{\label{fig:Close-up-images-of}Close-up images of the plume shown in Figure \ref{fig:171-Context}, at 2016 July 2 19:35UT in three AIA channels as marked, revealing the relative brightness and detail visible in each channel. These images show L1 data that have been PSF-corrected but otherwise unmodified.}
\end{figure}

\begin{figure}[tbh]
\begin{centering}
\includegraphics[width=6.25in]{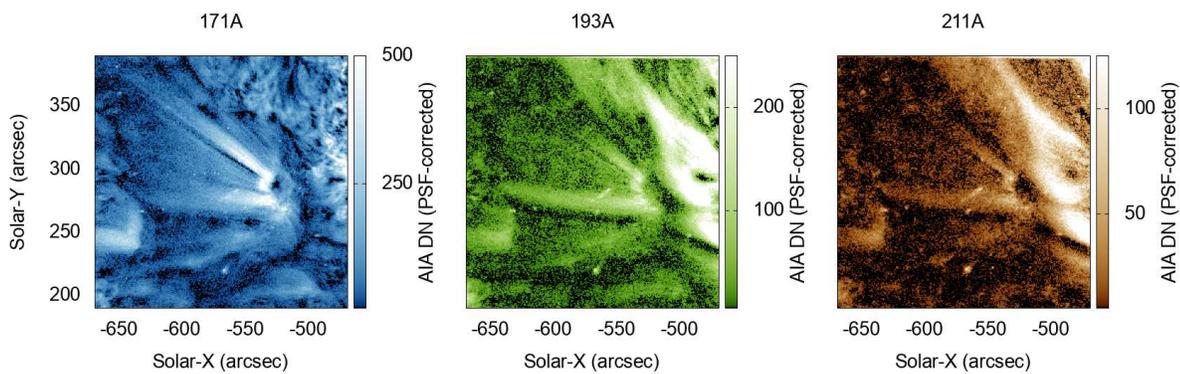}
\par\end{centering}
\caption{\label{fig:Unsharp-masked-images-of}Unsharp-masked images of the
plume shown in Figure \ref{fig:Close-up-images-of} reveal both fine
structure and excessive photon shot noise at the scales of interest
for this study.}
\end{figure}

\begin{figure}[tbh]
\begin{centering}
\includegraphics[width=6.25in]{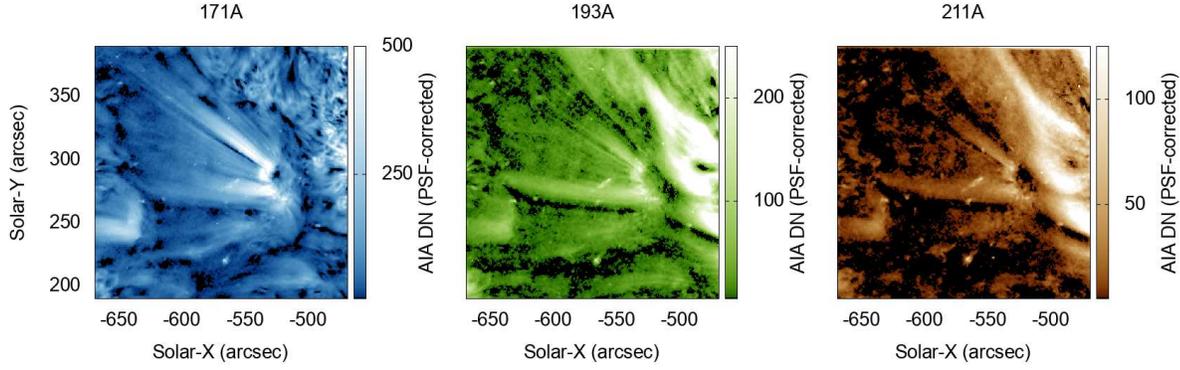}
\par\end{centering}
\caption{\label{fig:Noise-gated-versions-of}Noise-gated versions of the same
panels shown in Figures \ref{fig:Close-up-images-of} and \ref{fig:Unsharp-masked-images-of}
reveal the importance of noise detection and removal for fine structure
analysis. }
\end{figure}

\begin{figure}[tbh]
\begin{centering}
\includegraphics[width=6in]{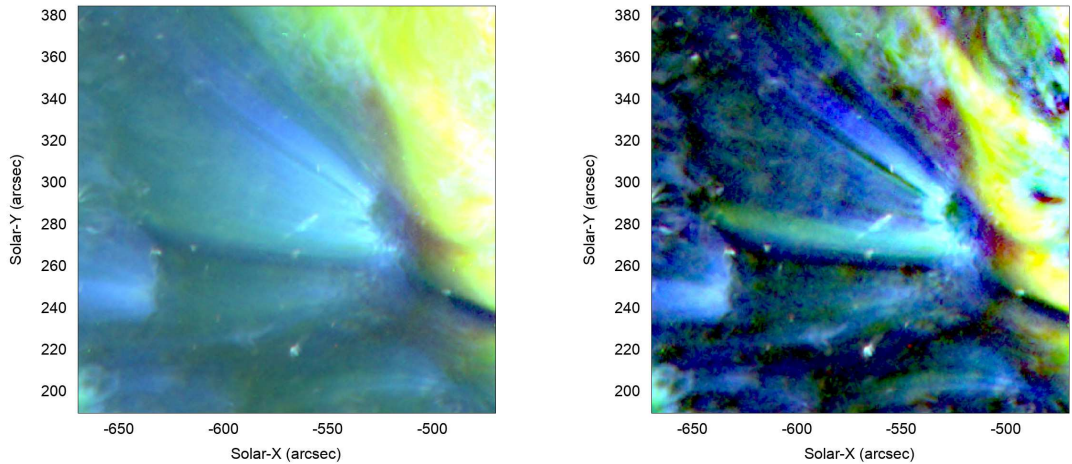}
\par\end{centering}
\caption{\label{fig:sRGB-composite-image}RGB composite image of the plume
shown in Figures \ref{fig:Close-up-images-of}-\ref{fig:Noise-gated-versions-of},
at the same time, reveals the ionization temperature structure at a glance.
Cooler (bluish) colors are dominated by the 171 \AA\ passband at or below 
$\sim 1$MK, while warmer (reddish) colors are dominated by the 211 \AA\ passband
at or above $\sim 2$ MK. Both direct transfer (left) and \emph{minsmooth}
unsharp masking with \emph{noise gating}  
(right) show multiple temperatures, but only the latter reveals the fine-scale structure in the plume. See \citet{uritsky_2020} for the animated version of the processed image set.}
\end{figure}

\item We processed the data to create the images using the \emph{noise-gating}
technique described by \citet{deforest_noise-gating_2017}. \emph{Noise gating} separates additive noise from an image sequence without imposing the
loss of resolution inherent in smooth kernel convolution. It treats
the image sequence as a 3D data set, and works by identifying features
that are coherent across space and/or time. Features are discriminated
from constant or image-dependent additive noise, based on the known
statistical and variational properties of the noise. Because AIA noise
levels are dominated by the Poisson statistics of photon counting
(``shot noise''), we used a gating threshold scaled with the image
brightness $B$ in each 12$\times$12$\times$12 pixel subregion, to
account for the $B^{1/2}$ dependence of shot noise. We set the gating
factor of the algorithm to $\gamma=3$. This process reduced the
photon shot noise by more than a factor of 10 in each AIA channel,
while preserving faint structures in the image sequence.

\item We unsharp-masked the \emph{noise-gated} data using the \emph{minsmooth} 
algorithm \citep{deforest_fading_2016}. \emph{Minsmooth} is an 
image operator that produces a smooth background model from an image by
finding the minimum (or low percentile) pixel value in the neighborhood
of each pixel in the image; the neighborhood size is set by an
aperture radius parameter. Unlike convolutional smoothing, \emph{minsmooth} 
produces an estimated minimum background model based on feature
scale. Subtracting this model yields an approximately positive-definite
image that contains only features smaller than the aperture used for
the operator. We applied \emph{minsmooth}  with an aperture radius of 20 AIA
pixels. The \emph{minsmooth}  unsharp masking highlights the spatially sharp
and/or variable components of the plume. Figures \ref{fig:Unsharp-masked-images-of} 
and \ref{fig:Noise-gated-versions-of} show the importance of the 
\emph{noise gating} step by contrasting simple
unsharp masking with \emph{minsmooth} (our step 3) with \emph{noise gating} followed
by unsharp masking (our steps 2 and 3), respectively. The \emph{noise 
gating} reduced shot noise by a factor of 10 or more, without reducing 
spatial or temporal resolution. This enables analysis of the temporal 
variations in the narrow, transient structures in this 
plume cluster.

\item We combined the 211\,\AA, 193\,\AA, and 171\,\AA~channels into a single
color movie sequence (Figure \ref{fig:sRGB-composite-image}) for further inspection 
and analysis, by placing each channel into the corresponding (red, green, blue) 
channel of an RGB triplet and encoding the triplet for display \citep{W3Cref}.  
\end{enumerate}

Figure \ref{fig:sRGB-composite-image} is a typical RGB composite image
from approximately 25,000 such composites 
during the 3.5-day interval under study. The relative line strengths
have been adjusted by their mean strengths over the entire disk, to
produce ``relative scaled'' color. Bluish
features emit mainly in the 171\,\AA~line (0.8 MK). Green and red
(or orange) features emit primarily in the hotter 193\,\AA~and 211\,\AA~lines,
respectively.  The color scheme immediately reproduces the familiar result 
that closed and quasi-closed structures (such as the quiet corona and 
streamer at right) are hotter than open coronal structures (such as the 
plume and coronal hole at left).

The left panel of Figure \ref{fig:sRGB-composite-image} shows direct, cleaned, extreme ultraviolet (EUV) radiances; the right panel uses \emph{ minsmooth}  unsharp masking to highlight small structures.  A broad plume (bluish) is visible near the center of the region of interest, extending to the left. The plume itself is readily seen to comprise both a broad, long-lived structure and narrow, more transient substructures (denoted 
``plumelets'' in this paper).An overall ``coronal haze'' pervades the region, even in the surrounding coronal hole. 

With any EUV images visible haze could be due to in-instrument scattering or real coronal effects \citep{deforest_2009}. These images were corrected using the validated, instrument-specific, inverse PSFs developed by \citet{poduval_point-spread_2013}, and hence have instrument scattering reduced by a factor of $\sim$30 compared to AIA Level 1 data.  We conclude that the observed haze in our processed frames is a real effect, not due primarily to instrumental scatter.

We analyzed these images to determine the origin, structure, and fluctuation characteristics of the plume, as described in \S \ref{sec:results}. 

The digital movie associated with Figure \ref{fig:Noise-gated-versions-of} reveals many small motions that are not apparent in the raw data.  We analyzed these motions using 2D time-series analysis \citep[e.g.,][]{deforest1998} as described in \ref{sec:dynamics}. The procedure relies on a time-distance representation of quickly propagating image features, in which it is essential that the sampling interval $\Delta t$ between the analyzed image frames be constant.

\begin{figure}[tbh]
\begin{centering}
\includegraphics[width=5in]{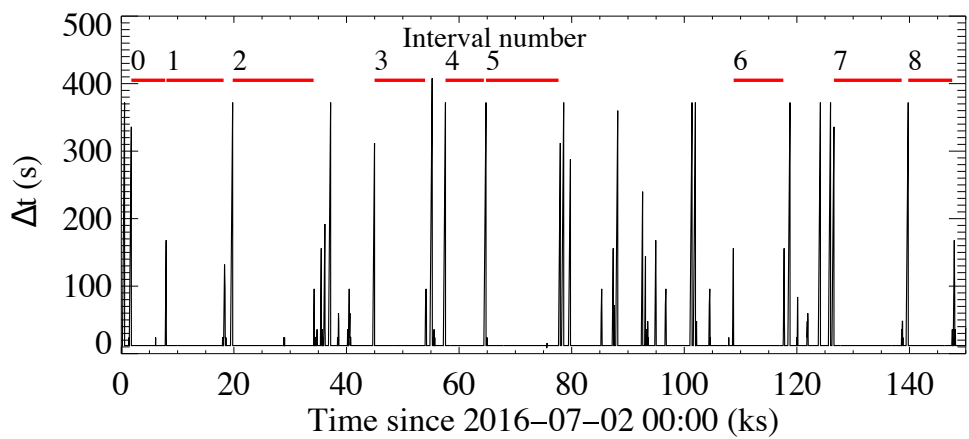}
\par\end{centering}
\caption{\label{fig:timeseries}Sampling history of AIA reveals nine time intervals with
sufficiently regular sampling for analysis of fluctuations in the plume and internal structure shown in Figure \ref{fig:sRGB-composite-image}. $\Delta t$ is the interval between successive images.}
\end{figure}

Figure \ref{fig:timeseries} shows that the complete data set was not fully continuous, with gaps as long as $\Delta t = 408$ s.  We selected several intervals of at least 500 images that were missing at most one image at 12-s cadence during the interval. For our spatiotemporal analysis, we identified nine such intervals in the analyzed image set (labeled 0 through 8 on Figure\ \ref{fig:timeseries} and in the following discussion). Since data gaps in the selected intervals accounted for less than $0.2\%$ of the interval duration, no time interpolation was necessary to fill the gaps.

Table \ref{table:intervals} provides information on the nine selected time intervals: the start and end date and time, the duration, and the number of AIA 171\,\AA~images collected.

\begin{deluxetable*}{ccccc}
\tablecaption{Time intervals selected for analysis\label{table:intervals}}
\tablewidth{0pt}
\tablehead{
\colhead{Interval} & \colhead{Start date and time} & \colhead{End date and time} & \colhead{Duration (s) } & \colhead{ Number of images} 
}
\decimalcolnumbers
\startdata
       0 & 2016-07-02 00:29:47 & 2016-07-02 02:09:59 &     6012 &      501 \\
       1 & 2016-07-02 02:13:35 & 2016-07-02 05:02:47 &    10152 &      844 \\
       2 & 2016-07-02 05:29:47 & 2016-07-02 09:29:23 &    14376 &     1196 \\
       3 & 2016-07-02 12:29:47 & 2016-07-02 14:59:23 &     8975 &      749 \\
       4 & 2016-07-02 15:59:47 & 2016-07-02 17:53:23 &     6816 &      569 \\
       5 & 2016-07-02 17:59:47 & 2016-07-02 21:34:23 &    12876 &     1073 \\
       6 & 2016-07-03 06:12:47 & 2016-07-03 08:39:47 &     8820 &      735 \\
       7 & 2016-07-03 11:09:47 & 2016-07-03 14:30:11 &    12024 &     1002 \\
       8 & 2016-07-03 14:49:47 & 2016-07-03 16:59:35 &     7788 &      650 \\
\enddata
\end{deluxetable*}

\section{\label{sec:results}Results}

\subsection{Plume structure}\label{sec:structure}

To characterize the small-scale filamentary structure in the plume, we 
enhanced the bright rays with the {\it Interactive Data Language} (IDL) version of the widely used Roberts cross edge-detection operator \citep{roberts1965}:
\begin{equation}
R_{m,n} = \left| L_{m,n} - L_{m+1,n+1} \right| + \left| L_{m,n+1} - L_{m+1,n} \right|,
\end{equation}
where $m$ and $n$ are the discrete coordinates of image pixels and $R$ is the Roberts
transform of the original image $L$. The first and the second terms on the right hand side 
are obtained from the image convolution using the masks
\begin{equation}
\begin{bmatrix}
0 & -1\\ 
1 & 0
\end{bmatrix},
\;\;
\begin{bmatrix}
1 & 0\\ 
0 & -1
\end{bmatrix}
\end{equation}
 and approximate the absolute value of luminosity gradient in the directions parallel and perpendicular to the principal diagonal of the $L$ matrix, respectively. The  edge-enhancement performance of the IDL routine is known to be the same as that of the original algorithm based on the root-mean-square value of the two gradient terms \citep{roberts1965}. We chose the Roberts operator for its robustness in the presence of high noise levels \citep{bonny2018}, which is essential for solar image processing. The resulting Roberts transform applied to AIA images is an uncalibrated proxy for the magnitude of the image-plane gradient of the squared column plasma density:
 \begin{equation}
     R(x,y) \propto  \left | \nabla \int n_e^2 d l \right |
 \end{equation}
 where $n_e$ is the volumetric electron number density and the integral is performed along the line of sight. 
  
\begin{figure}[tbh]
\begin{centering}
\includegraphics[width=6.5in]{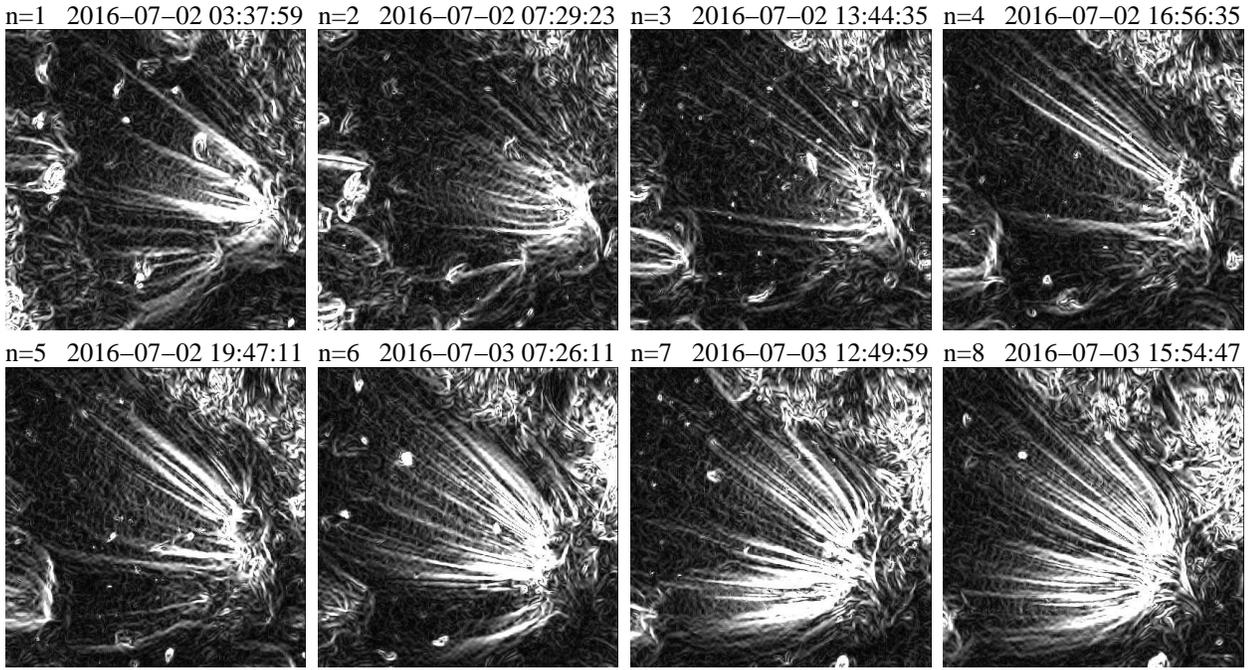}
\par\end{centering}
\caption{\label{fig:roberts_transform} Roberts-transformed AIA 171\,\AA~images of the plume for the  gap-free time intervals shown in Figure \ref{fig:timeseries}, excluding Interval 0 which was similar to Interval 1. A complex, time-evolving structure is evident.}
\end{figure}

 Figure \ref{fig:roberts_transform} shows Roberts transforms of AIA 171\,\AA~ images taken in the middle of eight contiguous time intervals (labeled with index $n$) identified in Figure \ref{fig:timeseries}. Interval 0 is very similar to Interval 1 and is not shown. The grayscale coding represents the $R$ range between 0 and 30 units; for comparison, the maximum value of $L$ is $\approx 1000$ units. A highly complex, filamentary structure of the interior region of the plume is evident. The thickness of the smallest $R$ features, presumably representing boundaries of the individual plume rays, are close to the image resolution; the largest features reach more than 10 pixels across. 
 
Another notable feature of the Roberts-transformed images is long-term systematic evolution of the plume structure over the entire period covered by the image set. Intervals 1 to 3, covering the first $\approx 15$ hours from the start on 2016 July 2, are characterized by a sparse and clustered ray pattern, with some of the plume sectors containing multiple features and others showing no such features. Interval 3 reveals the smallest number of identifiable ray edges. After that, an increasingly  refined and uniform ray structure gradually develops, reaching its mature state by Interval 8 during which rays permeate nearly the entire plume. Later, the fine-scale structure becomes less pronounced, which could in part be explained by changes in the plume orientation relative to the line of sight. 

\begin{figure}[tbh]
\begin{centering}
\includegraphics[width=4.8in]{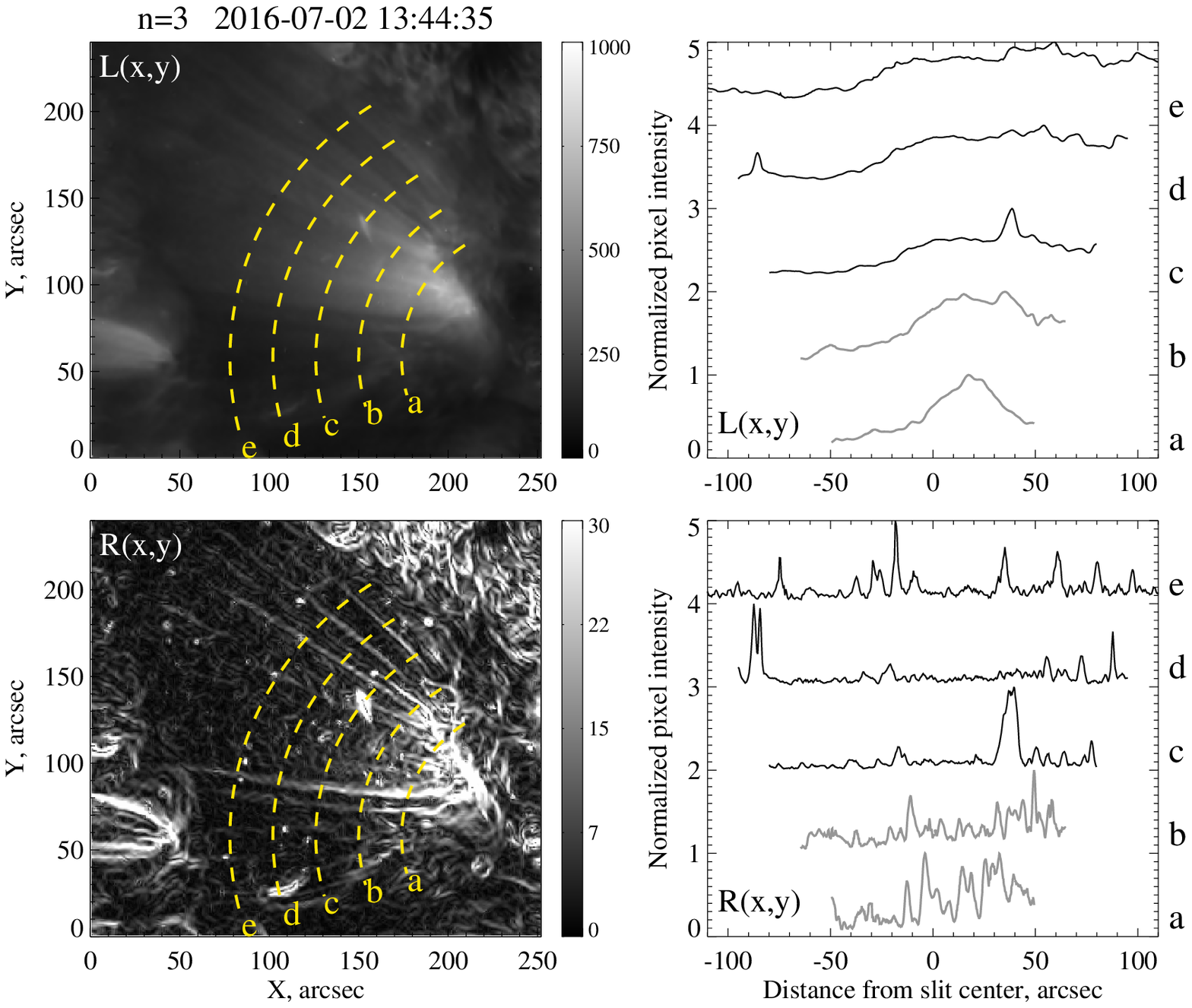}
\vskip 0.1in
\includegraphics[width=4.8in]{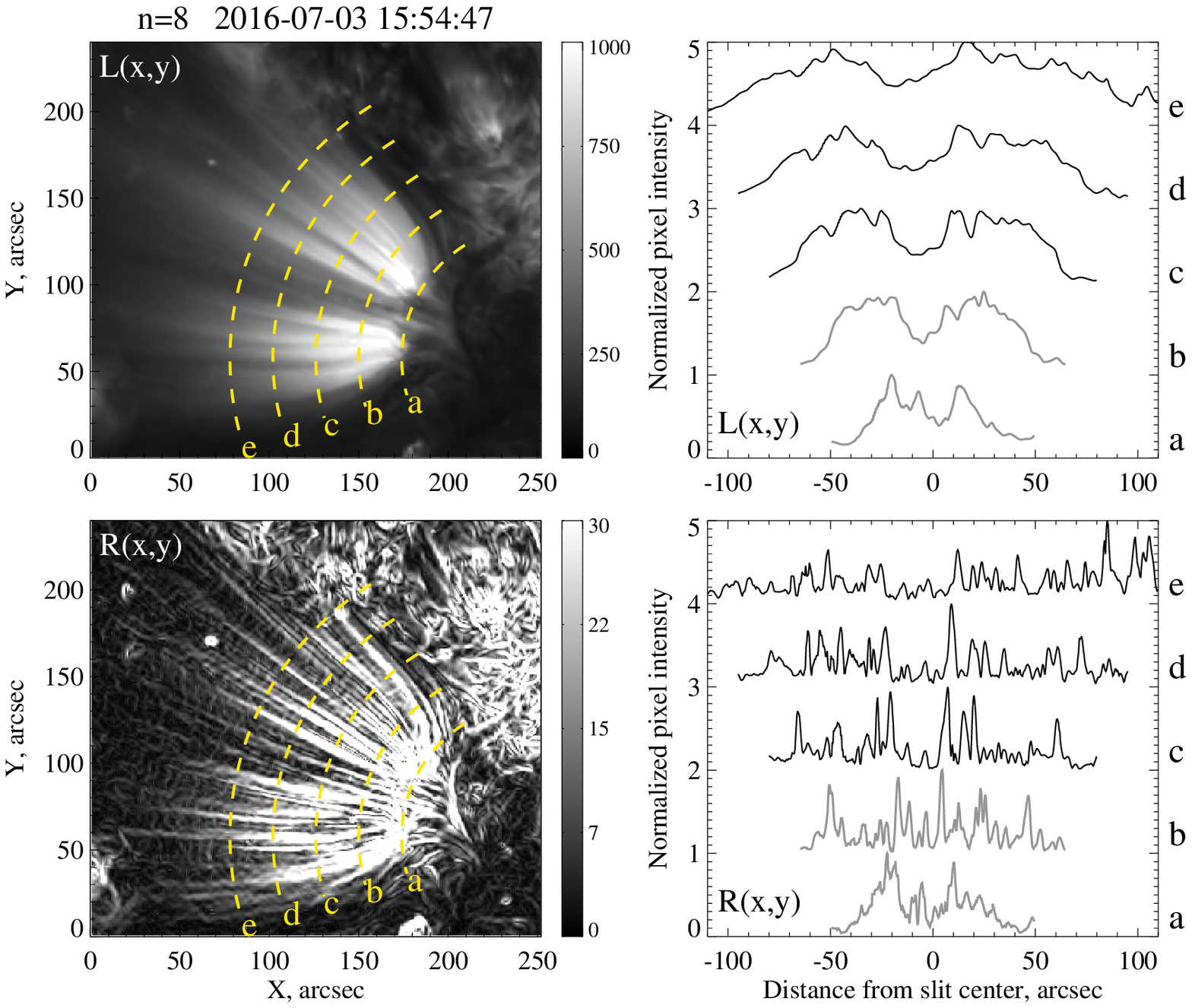}
\par\end{centering}
\caption{\label{fig:roberts_slits}Left: Zoomed-in AIA 171\,\AA~preprocessed ($L(x,y)$) and Roberts-transformed ($R(x,y)$) plume images during Intervals 3 and 8 representing the early and late plume evolution, respectively. Right: Normalized intensity of the same images along the five arc-shaped slits shown with dashed yellow lines on the left panels. Both intervals exhibit a well-developed filamentary structure near the plume base, with Interval 8 showing significantly more small-scale ray edges at higher altitudes.}
\end{figure}

We constructed 1D profiles of the $R$ maps along a set of arc-shaped virtual slits, labeled from $a$ to $e$ in Figure \ref{fig:roberts_slits}, with the center of curvature chosen to ensure that most of the slits cross the plumelet edges at an approximately 90$^\circ$ angle. Figure \ref{fig:roberts_slits} shows five such profiles for time Intervals 3 and 8 exhibiting, respectively, the least and most developed ray structure. To reduce the noise level, the profiles were averaged over $\pm 5$ pixels across the slits, normalized by their maximum values, and shifted along the vertical axis for easier comparison. Signals from the slits labeled $a$ and $b$, shown in gray, are less reliable because they were contaminated by the underlying coronal moss seen through the optically thin plume material. 

Compared to the earlier Interval 3, Interval 8 shows much more structure in the 1D profiles, which reveals a multitude of small-scale rays over the entire length of the slits (Figure\ \ref{fig:roberts_slits}). The transverse scale of the plumelets (i.e., the distance between adjacent peaks in $R$) during Interval 8 ranges from $\sim 1''$  to $\sim 10''$; depending on the detection approach, one can identify 30-35 individual features in slits $c$-$e$, thus yielding a characteristic transverse scale of 5-6$''$. 

\begin{figure}[tbh]
\begin{centering}
\includegraphics[width=3.5in]{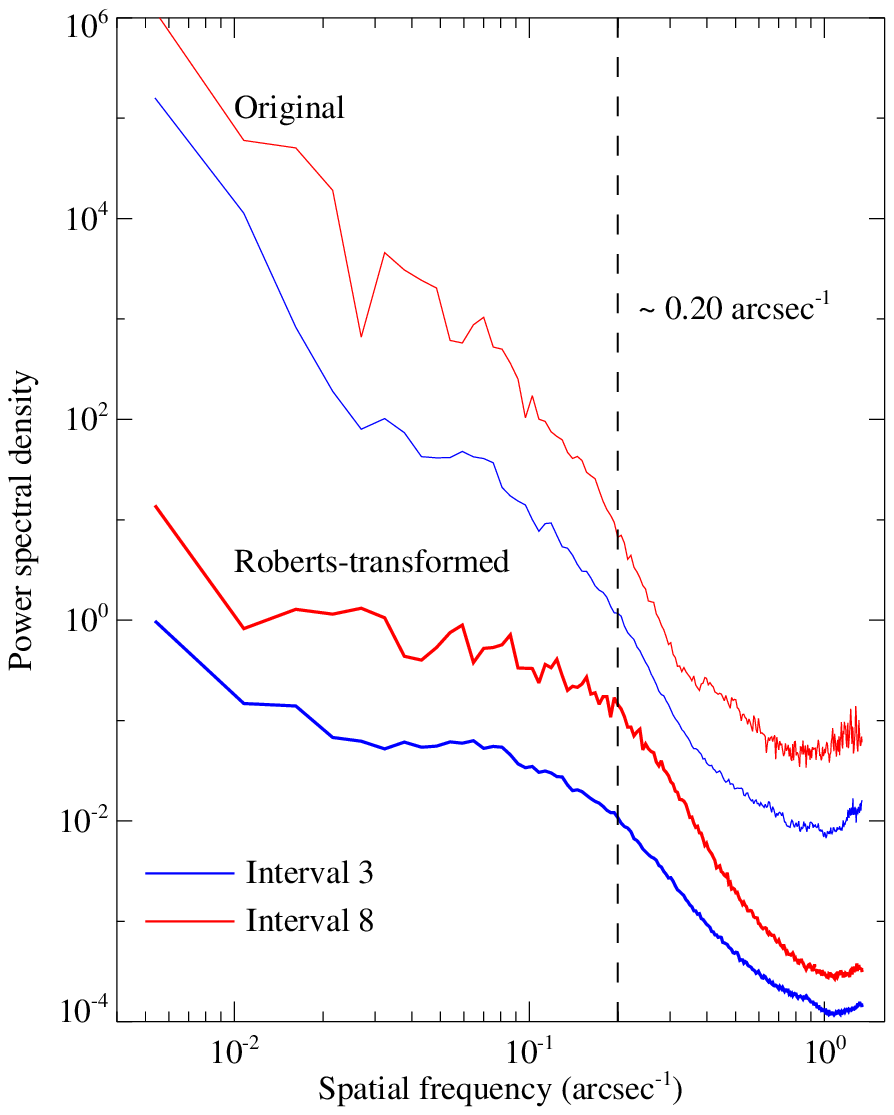}
\par\end{centering}
\caption{\label{fig:spatial_spectra} Fourier power spectra of AIA 171\,\AA~intensity fluctuations across the plume for Intervals 3 and 8 before and after applying the Roberts transform. The spectra were averaged over multiple time steps, and over 11 arc-shaped slits positioned between the slits labeled ``c'' and ``e'' in Figure \ref{fig:roberts_slits} to avoid moss contamination near the plume base. The spectra of the original images (thin lines) are shifted up by two decades for easier comparison with the spectra of Roberts-transformed images (thick lines) used to estimate the characteristic spatial scale.}
\end{figure}

This scale estimate was verified by a Fourier analysis of an extended set of arc-shaped profiles averaged over multiple time steps. The region between slits $c$ and $e$ in Figure \ref{fig:roberts_slits} was resampled by 11 equally spaced slits that were used to construct 11 1D profiles, smoothed over $\pm 3$ pixels across the slit. For each profile, we computed a spatial power spectrum using a Hanning-filtered Fast Fourier Transform. The procedure was repeated for every fifth image collected during the specified time interval, after which the power spectra obtained for the individual image slits and time steps were averaged:

\begin{equation}
    \bar{P}(k) = \left\langle  P_{m,n}(k)\right\rangle_{m,n}.
\end{equation}
Here, $P_{m,n}$ is the local power spectrum computed for the $m$th slit ($m=1,...,11$) of the $n$th image, $k$ is the spatial frequency, and $\bar{P}$ is the averaged power spectrum.

Figure \ref{fig:spatial_spectra} shows two pairs of the slit- and time-averaged spatial power spectra $\bar{P}$ for the original ($L(x,y)$) and the Roberts-transformed ($R(x,y)$) images during Intervals 3 and 8. The spectra computed using the original images show no identifiable small-scale structure across the plume, consistent with the very small dynamic range of the rays in the $L(x,y)$ images compared to the background signal. In contrast, the spectra of the edge-enhanced images have well-defined characteristic scales that are different for the two intervals. These scales are manifested in the form of a spectral ``break'' separating frequency ranges described by small and large log-log slopes. This type of spectral signature is indicative of a stochastic spatial pattern, as in a random telegraph signal for example, as opposed to isolated spectral peaks produced by periodic patterns. The spectral break characterizing Roberts-transformed plume images in Interval 8 is located near the spatial frequency $\approx 0.2$ arcsec$^{-1}$, again yielding a transverse scale of about 5$''$. This spectral estimate is consistent with the shape of the profiles in Figure \ref{fig:roberts_slits}.

Scale measurements based on the edge-enhanced $R$ images should be interpreted with care. Because $R(x,y)$ is an unsigned gradient-based measure, the characteristic sizes of the structures in the original $L(x,y)$ images should be about two times larger that those in the transform. Thus, for instance, a periodic structure $L \propto \sin (k_\perp s) $ in the direction $s$ perpendicular to the plume's main axis would be transformed into the structure $R \propto [1 + \cos (2 k_{\perp}s) ]^{1/2}$, described by the wavelength $\pi/k_{\perp}$ that is half that of the structure in the original image. 

Therefore, the actual ray scale is about twice the spatial separation between the 
ray edges derived from the Roberts-transformed images (Figure\ \ref{fig:roberts_slits}) or 
the characteristic spatial scale derived from the spectral analysis (Figure\ \ref{fig:spatial_spectra}). The estimated transverse Roberts scale of 5-6$''$ therefore corresponds to a characteristic ray scale of 10-12$''$.

\begin{figure}
\begin{centering}
\includegraphics[width=3.5in]{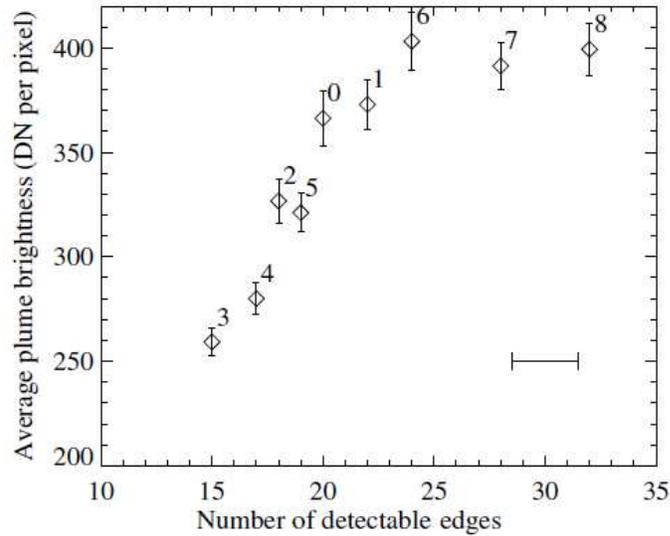}
\par\end{centering}
\caption{\label{fig:intensity_filamentation} Average brightness of the plume (relative units) during the first nine time intervals (Intervals 0 through 8) shown in Figure \ref{fig:timeseries}, versus the approximate number of plumelet edges identifiable in the Roberts-transformed images of the plume. Vertical error bars represent standard errors 
of average brightness measured in different longitudinal slices (see Figure \ref{fig:grids}); the horizontal error bar shows a typical uncertainty in counting the edges. }
\end{figure}

The average intensity of the 171\,\AA~emission from the entire plume region (as defined by
the outer boundary of the adjustable curvilinear grid described in the next section) versus the number of detectable plumelets on the edge-enhanced Roberts-transformed images is plotted in Figure \ref{fig:intensity_filamentation}. As mentioned above, the filamentary structure of the plume evolves slowly through two phases. During Intervals 0 and 1, a moderate number of edges produce a plume of nearly peak brightness. Intervals 1 through 6 manifest a decrease followed by an increase in both the number of ray boundaries identified in the Roberts-transformed images and the average plume brightness. Finally, the plume reaches a saturation stage during Intervals 6 to 8 when an increase in the number of plumelets is no longer accompanied by an average brightness increase. Because we used stray-light-deconvolved AIA data, the observed dependence of the midscale diffuse brightness on the number of small-scale bright features is unlikely to be attributable to PSF effects in the EUV data. Regardless of the temporal sequence, Figure \ref{fig:intensity_filamentation} clearly shows that the two plume properties are strongly correlated throughout the early evolution of the plume structure. While concurrent evolution of two empirical parameters does not necessarily imply a casual connection, the correlation between the amount of plume filamentation and brightness demonstrated in Figure \ref{fig:intensity_filamentation} suggests that well-developed fine structure may be required to form an optically intense plume.

Several alternative edge-enhancement methods, including the Sobel-Feldman operator and a simple shift-difference filter, were also tested (the results are not shown). The plumelet structures identified with these additional methods were found to be nearly indistinguishable from those obtained using the Roberts transform, confirming that the detected fine-scale structure of the plume is independent of the particular edge-enhancement algorithm used and represents an objective physical phenomenon.

\subsection{Plume evolution}\label{sec:dynamics}

\begin{figure}[tbh]
\begin{centering}
\includegraphics[width=6.5in]{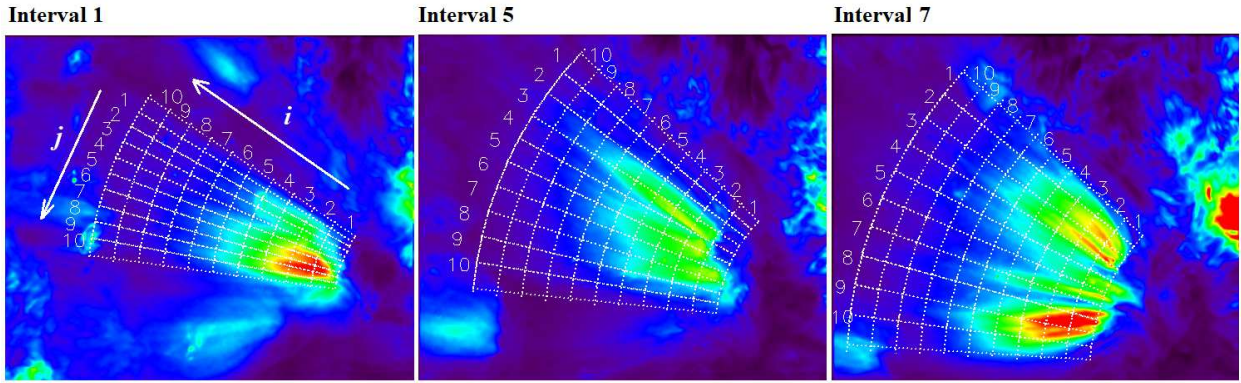}
\par\end{centering}
\caption{\label{fig:grids} Examples of adjustable curvilinear grids used to construct time-distance plots in longitudinal and transverse directions relative to the plume. The background snapshots are AIA 171 \AA ~images in the middle of each time interval, preprocessed as described in \S\ref{sec:data}. The dark blue (bright red) colors of the chosen rainbow palette correspond to 0 (1000) AIA DN units.}
\end{figure}

To characterize the longitudinal and transverse dynamics of the plume rays in each interval from Figure \ref{fig:timeseries} systematically, we constructed by hand a fan-like feature-matched coordinate system, following \citet{uritsky13}.  The coordinate system is overlaid on the plume during Intervals 1, 5, and 7 in Figure \ref{fig:grids}.  The feature-matched coordinates $x'$ and $y'$ run across and along the visually identified "grain" of the plume in the image plane.  We divided the $x'$ and $y'$ axes into 10 windows numbered 1 to 10, as shown in Figure \ref{fig:grids}, each covering an identical range of polar angles.

To visualize the propagating disturbances and structures present in each of the slices defined by the feature-matched grid, we averaged the pixel brightness over $y'$ for each transverse slice, which resulted in 10 transverse time-distance plots. We also averaged pixel brightness values over $x'$ for the 10 longitudinal slices, to obtain time-distance plots reflecting the dynamics along the plume. 

\begin{figure}[tbh]
\begin{centering}
Slits across plume \hspace{1.5in} Slits along plume
\includegraphics[width=5.2in]{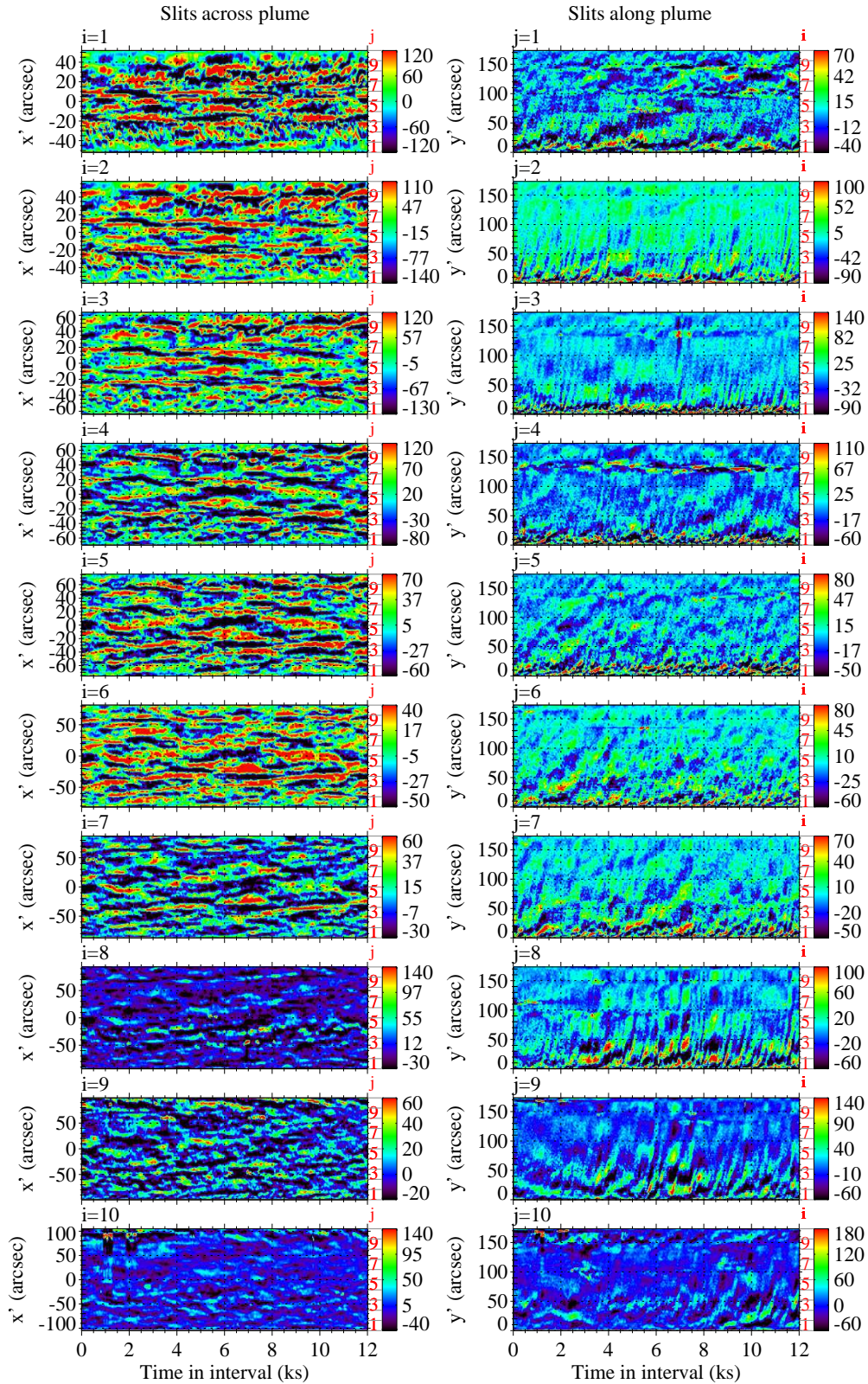}
\par\end{centering}
\caption{\label{fig:interval_7} Time-distance plots for Interval 7,
constructed by averaging the AIA 171\,\AA~intensity along transverse (left) and longitudinal (right) virtual slits defined in Figure \ref{fig:grids}. The plots were detrended in order to enhance transient features.}
\end{figure}

Figure \ref{fig:interval_7} shows an example of time-distance plots in the longitudinal and transverse directions relative to the main plume axis during Interval 7. To extract significant features above the background emission, the plots were detrended in the spatial and temporal directions by subtracting respectively 5th and 3rd order polynomials fitted to the data, following a previously tested methodology \citep{uritsky13}. The residual fluctuations comprise between 9 and 25$\%$ of the total emission in the slits. 

\subsubsection{Transverse structure}\label{sec:transverse}

In the transverse time-distance plots (left panels in Figure\ \ref{fig:interval_7}), the narrow elongated features aligned with the horizontal axis are the evolving signatures of the rays discussed in \S\ref{sec:structure}. The number of features per plume cross-section varies significantly with slit location and the observation time. Slits with indices 
$i=3-7$ show the most pronounced filamentary structure, with a typical number 
of identifiable rays ranging between 9 and 13. Considering the average size of these slits ($\approx 120''$), the characteristic scale of these features should lie between $9''$ and $13''$. This agrees well with the estimated widths based on our analysis of 1D profiles from the Roberts-transformed images (\S\ref{sec:structure}). The slits with $i=1,2$, located near the plume base, contain fewer detectable rays (only 7 to 8), possibly due to a stronger projection effect resulting in multiple overlapping rays. The slit with $i=10$, farthest from the plume base, has the weakest detrended signal, making it difficult to identify any features.

The positions of the rays on the transverse time-distance plots are relatively stable; however, their brightness seems to vary slowly over a time scale of about 5000-6000 s (80-100 min). This period cannot be reliably measured because it is comparable to the duration of the entire observation. In addition, it may be affected by the polynomial detrending applied to the plots. A close inspection of the time-distance plots also reveals a quasi-periodic, checkboard-like pattern in the maxima and minima of the detrended brightness, indicating that the emission crests and troughs at neighboring locations across the plume are out of phase.

\subsubsection{Longitudinal structure}\label{sec:longitudinal}

The longitudinal time-distance plots for Interval 7 (right panels in Figure\ \ref{fig:interval_7}) reveal an intense, consistent pattern of quasi-periodic, upwardly propagating disturbances. In \S\ref{sec:discussion} we discuss whether these disturbances are bulk motions, waves, or a combination. The characteristic time lapse $T$ between the features ranges from $T \approx 200$ s in the longitudinal slits with indices $j=2,3$ to $T \approx 750$ s in slits with $j=6,7$. The slits indexed $j=8,9$ exhibit the most stable and intense  observed over the entire slit length, with period $T \approx 200$-$400$ s as estimated by a visual inspection of the time-distance plots. We derive more precise estimates of $T$ from a quantitative analysis of these plots in \S \ref{sec:longdist}. 

\begin{figure}[tbh]
\begin{centering}

\includegraphics[width=3.0in]{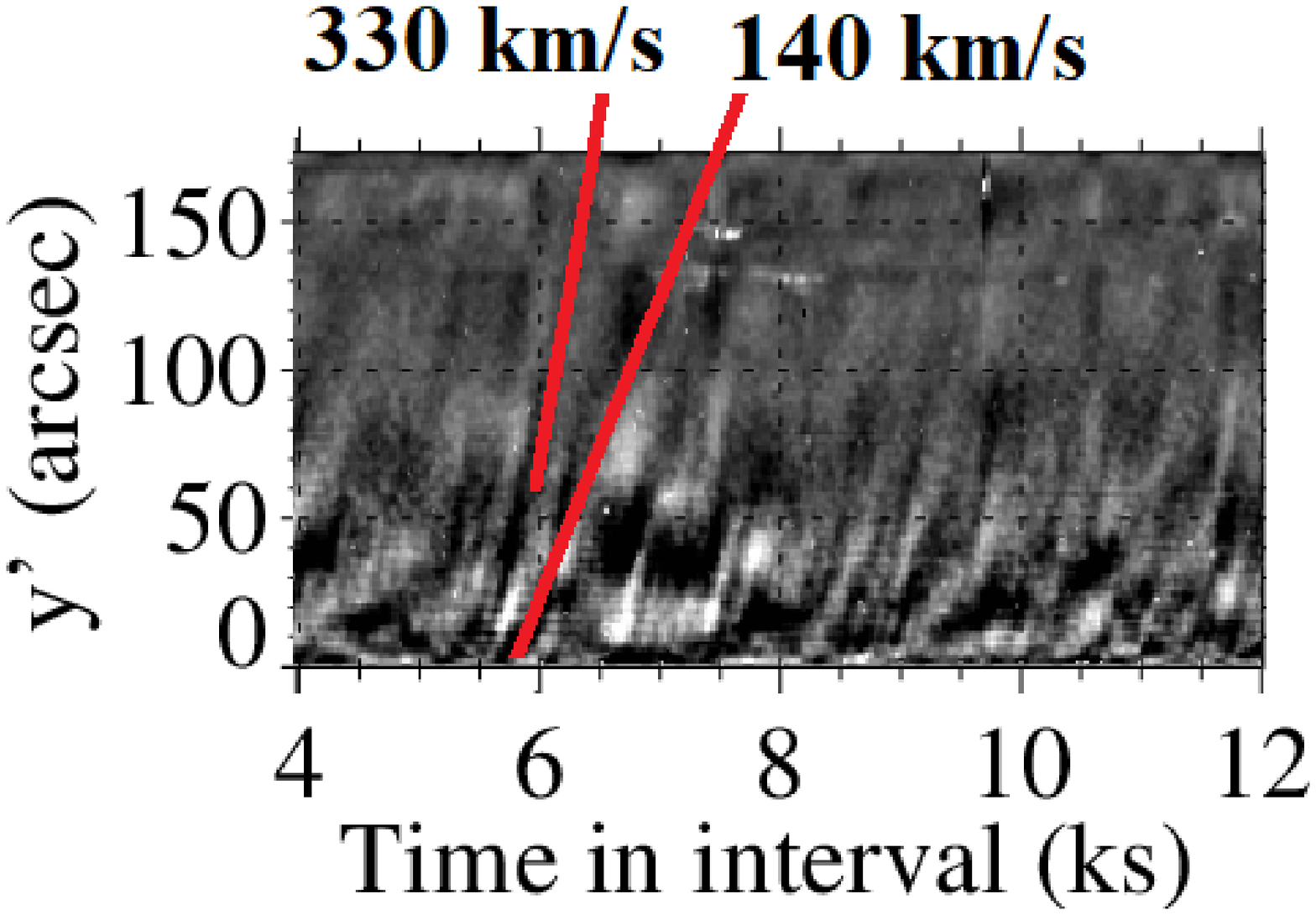}

\includegraphics[width=2.0in]{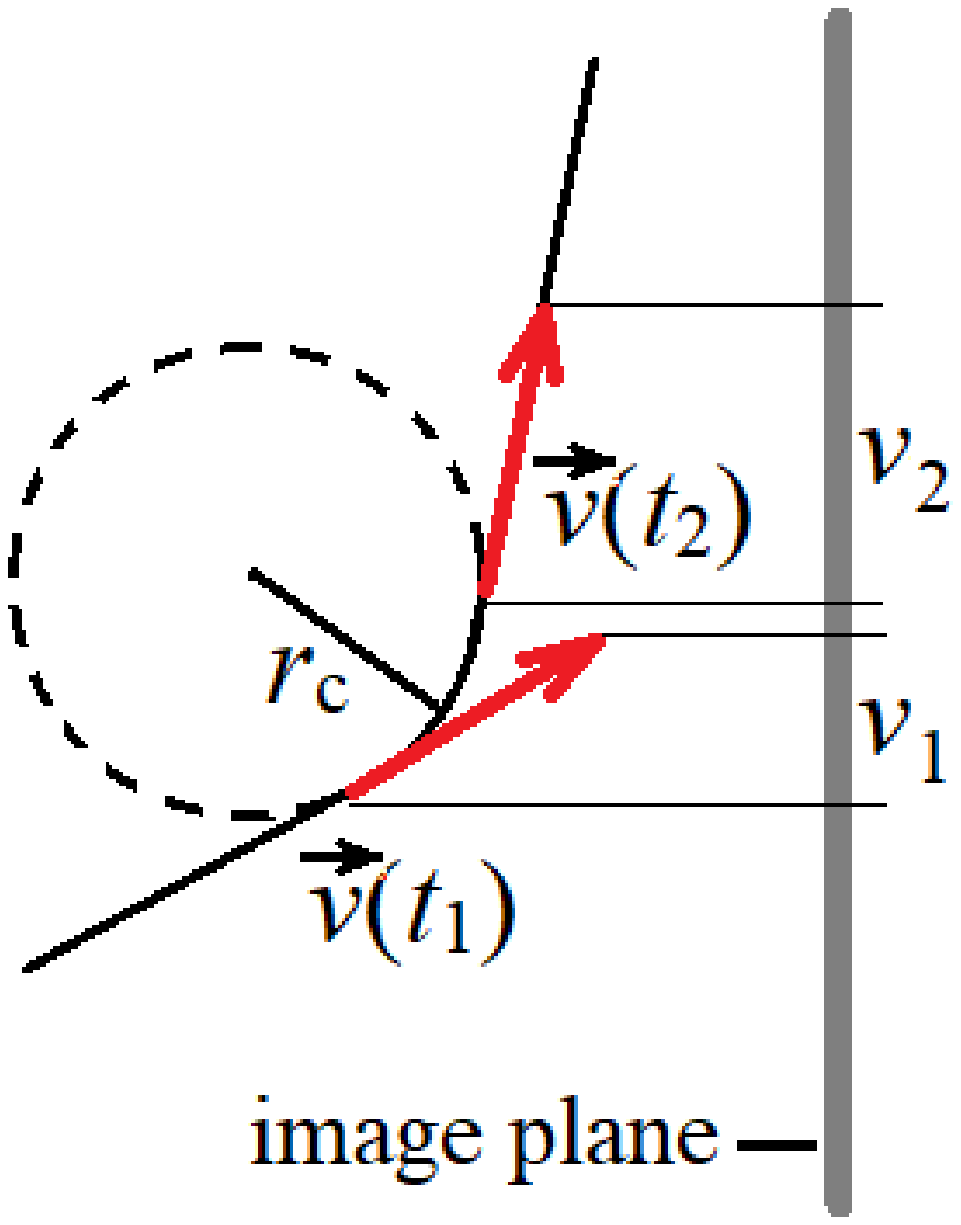}
\par\end{centering}
\caption{\label{fig:interval_7_j_9} Top: A zoomed-in time-distance plot in the longitudinal slit $j=9$ of Interval 7 showing steepening of propagating fronts with altitude ($y'$). The red lines on the top panel show the slopes of one bright front measured at two different heights. Bottom: A simplified vertical cut through the plume base, where the magnetic field expands rapidly before becoming more collimated, illustrating the propagation geometry assumed in Eq.\ \ref{eq:curvature}. The solid black curve represents a section of a flux tube defining the low plumelet geometry. The red arrows illustrate the expected change of the velocity direction due to the plume base curvature, corresponding to the change in slope shown above.}
\end{figure}

The slits showing consistent dynamic activity across a wide range of projected altitudes manifest a clear tendency for the phase speed of the propagating front to increase with distance from the plume base. An example of such dispersion is shown in the top panel of Figure \ref{fig:interval_7_j_9}, which provides a zoomed-in view of the longitudinal time-distance plot in slice $j=9$ during Interval 7. The front highlighted with two straight line segments, indicating its approximate local slopes, is significantly steeper at high altitudes than at low altitudes; its speed apparently increases from about 140 km s$^{-1}$ near the plume base to about 330 km s$^{-1}$ in the upper half of the slit. A similar acceleration versus height can be seen in other fronts. The observed speed increase likely represents a partial rotation of the velocity vector caused by the curved shape of the magnetic flux tubes near the base of the plume, as illustrated in Figure \ref{fig:interval_7_j_9} (bottom). By approximating the curved portion of the plume with a circular arc, the apparent speed change can be related to the angular velocity and thus to the geometry of the plume:

\begin{equation}
    \left | \frac{\cos^{-1}(v_2/v) - \cos^{-1}(v_1/v) }{t_2 - t_1}\right | \approx \frac{v}{r_c} ,
    \label{eq:curvature}
\end{equation}
where $v_1$ and $v_2$ are the image plane projections of the phase speed observed respectively before and after the front entered the curved path, $t_1$ and $t_2$ are the corresponding times, $r_c$ is the radius of curvature of the lower portion of the plume, and $v$ is the true speed of the front. 

We solved Equation \ref{eq:curvature} numerically to estimate $v$. By substituting $v_1 = 140$ km s$^{-1}$, $v_2 = 330$ km s$^{-1}$, and $t_2-t_1 \approx 250$ s for the front featured in Figure \ref{fig:interval_7_j_9} (top) and $r_c = 160$ Mm for the average curvature radius as suggested by shape of the plumelet edges (see Figure \ref{fig:roberts_transform}), one obtains $v \approx 400$ km s$^{-1}$. This speed is greater than $v_2$ because of projection effects. The corresponding plume angle characterizing its approximate orientation with respect to the image is about $34^\circ$. This angle is substantially smaller than  $\cos^{-1}(\sqrt{\bar{x}^2 + \bar{y}^2}/R_S) \approx 62^\circ$ representing the local normal to the Sun's surface, where $\bar{x}$ and $\bar{y}$ are the average plume coordinates in the image plane and $R_S$ is the solar radius. Therefore the plume is not radially oriented but is tilted away from the local normal by about $30^\circ$.

\subsubsection{Longitudinal fluctuations}\label{sec:longdist}

\begin{figure}[tbh]
\begin{centering}
\includegraphics[height=8in]{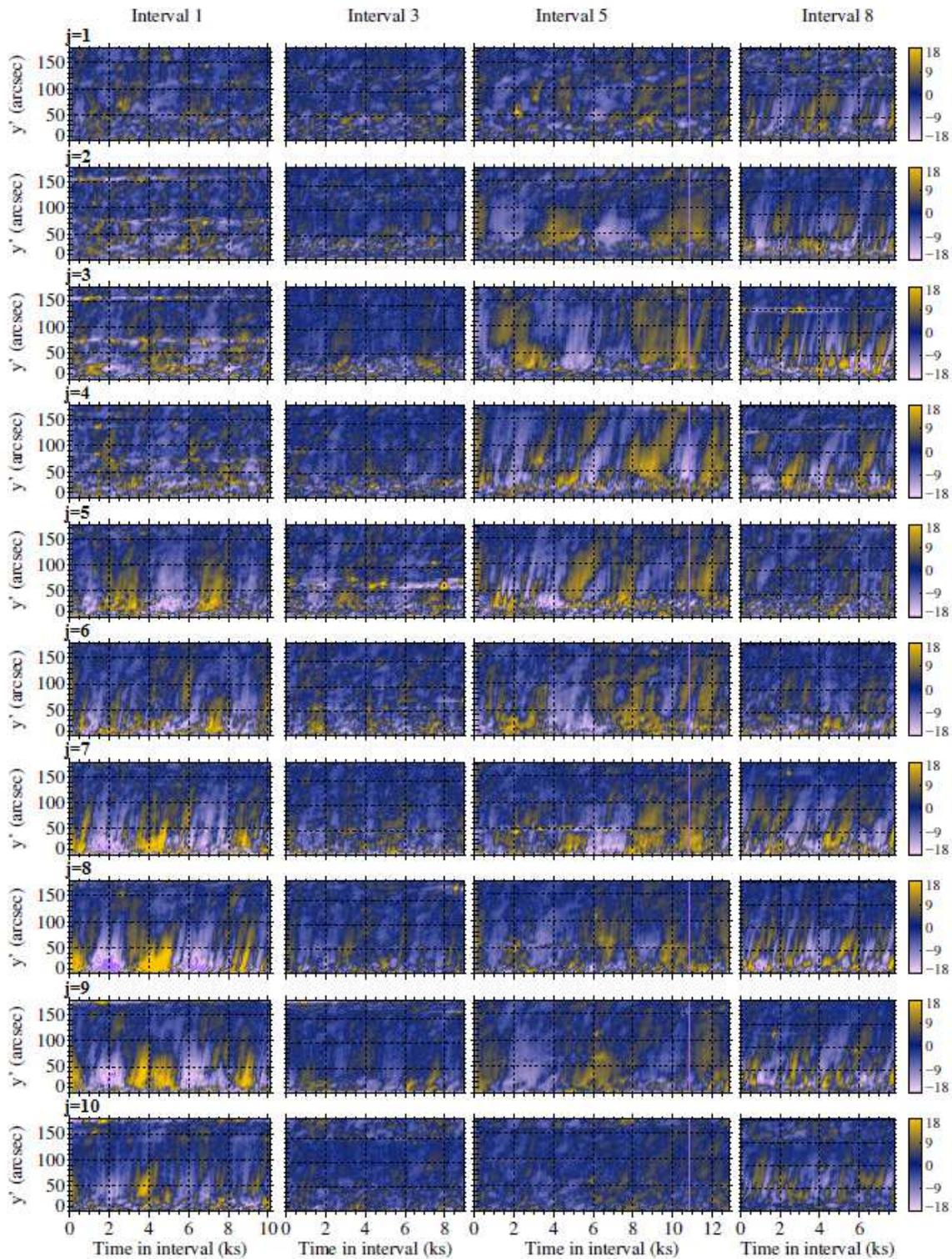}
\par\end{centering}
\caption{\label{fig:four_intervals} Time-distance plots for time Intervals 1, 3, 5, and 8 constructed by averaging the AIA 171\,\AA~intensity along longitudinal virtual slits defined in Figure \ref{fig:grids}. The plots highlight the longitudinally propagating disturbances and capture the differences between the rising and mature phases of the plume development. The disturbances show the most consistent periodic pattern across all plume sectors during Interval 8, when the plume was extremely filamented, whereas the pattern is barely detectable during Interval 3, when the plume contained only a few substructures (see Figure\ \ref{fig:roberts_transform} for reference). Intervals 1 and 5 exhibit propagating disturbances inside the most filamented sectors of the plume. }
\end{figure}

To extract longitudinal dynamics from the data, we produced detrended charts of the evolution along specific locations in the plume.  To ensure that we captured spatial variations, we detrended only in time:  each pixel location in the flat-fielded, tracked movie was treated as a single time series.  We modeled each time series with a cubic polynomial and subtracted that polynomial from the data.  This yielded absolute-brightness fluctuations in intensity units of corrected AIA DN. The resulting plots, with a common color table and dynamic range centered on 0, are displayed in Figure  \ref{fig:four_intervals} for four time intervals. 

The columns in Figure \ref{fig:four_intervals} reflect different stages of long-term plume evolution, from the weak, sparse filamentary structure in Interval 3 to the well-developed structure evident in Interval 8. Multiple time scales and multiple speeds co-exist in single panels. For example, during Interval 5 at $j$=5, disturbances lasting of order 1-2 ks co-exist with others lasting only about 100 s. The disturbances at the two time scales propagate at speeds that differ by about a factor of two, as illustrated in Figure \ref{fig:interval_7_j_9}. Both time scales are present throughout the data set, although the relative amplitude changes across intervals and across $j$ values within each interval, with the more intense high-frequency component typically present near the base.

\begin{figure}[tbh]
\begin{centering}
\includegraphics[width=4.5in]{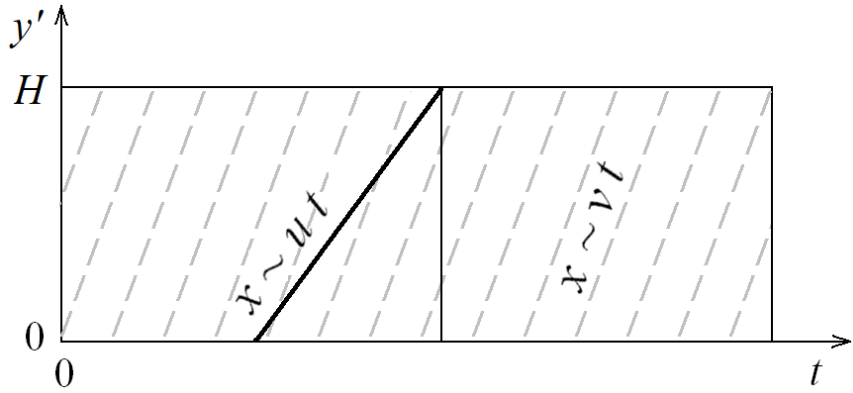}
\par\end{centering}
\caption{\label{fig:surfing} Schematic drawing \citep[adapted from][]{uritsky13} illustrating the extraction of the surfing averaged signal from a time-distance plot representing propagating disturbances in a longitudinal slice of the plume. Dashed lines show the fronts of a periodic disturbance moving with phase speed $v$. The solid line is an example of an averaging path defined by the surfing speed $u$. }
\end{figure}

To further investigate the short-term time variability of the plumelets caused by the propagating disturbances, we used an adaptive numerical technique \citep{uritsky09, keiling12, uritsky13} designed to identify wave signals in time-distance plots with low signal-to-noise ratios, such as the plots in Figure \ref{fig:four_intervals}. The technique is based on analysis of the velocity-dependent ``surfing average'' signal $S(t,u)$  illustrated in Figure \ref{fig:surfing}. The surfing average is defined as a time-dependent mean of the time-distance plot $\hat{L}(y',t)$ computed along a set of straight parallel world lines running through the starting position $y'=0$ at different times, with a fixed slope $u$ representing an assumed phase speed:

\begin{equation} 
S(t,u) = \frac{u}{H} \int\limits_t^{t+ H/u } \hat{L}(y'=y_S, t') dt'.
\label{eq:surfing}
\end{equation}

Here, $t$ is the running time, $y_S = u\,(t' - t)$ is the spatiotemporal averaging path, and $H$ is the propagated distance in the image frame. 

If $\hat{L}(y',t)$ contains a periodic disturbance propagating with speed $v$, then the dynamic range of $S(t,u)$ maximizes at $u = v$ since in that case some of the surfing lines will be sitting on the troughs and the crests of the wave or flow. For a sufficiently strong and consistent signal, maximizing the value of $S$ allows for an automatic evaluation of $v$ \citep{uritsky13}. We were unable to invoke this approach here because the activity in many of the studied time intervals was irregular and weak. Instead, we detected the fronts in the detrended time-distance plots by eye. For each time interval, we selected the longitudinal slice in which the propagating disturbance was the most pronounced in terms of its amplitude, temporal stability, and  spatial extent. The identified fronts were used to calculate the average characteristic value of the phase speed $v$ based on the maximization of the dynamic range of the surfing signal (Eq. \ref{eq:surfing}). The period $T$ of the propagating disturbances was estimated from the average time delay within adjacent pairs of successfully detected wave fronts as well as from the position of the main peak on the slit - averaged Fourier spectrum.

\begin{figure}[tbh]
\begin{centering}
Interval 3 \hspace{2.0in} Interval 8
\includegraphics[width=2.7in]{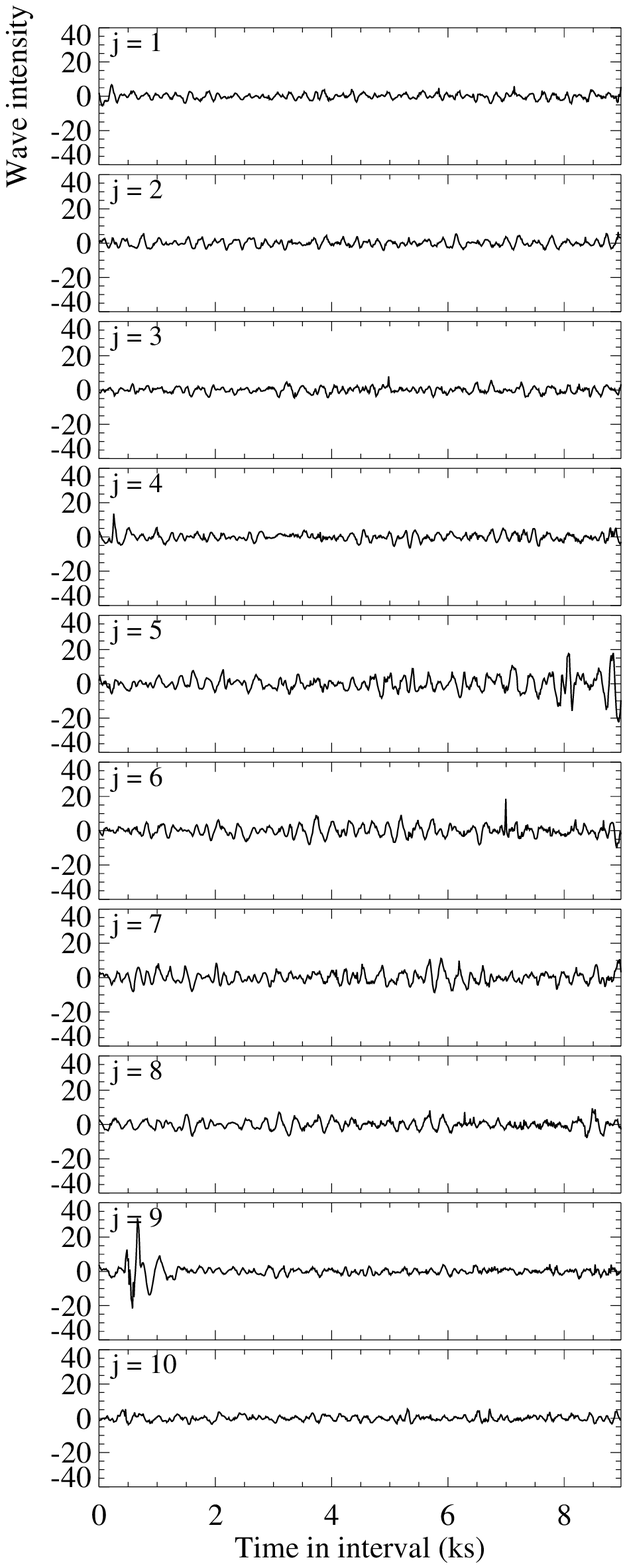}\includegraphics[width=2.7in]{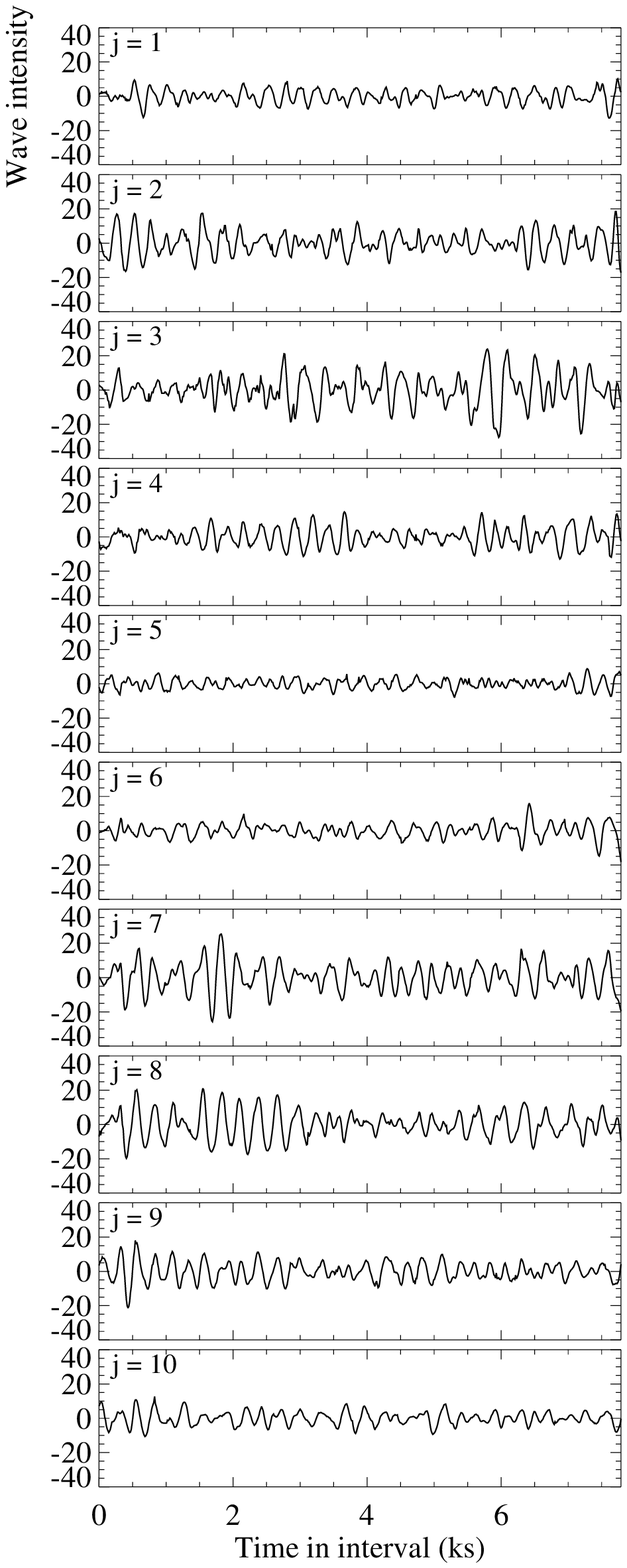}
\par\end{centering}
\caption{\label{fig:waveforms} Propagating-front signals $S(t,u)$ obtained using the surfing average technique for observed Intervals 3 (left) and 8 (right) in each of the 10 longitudinal slits. Note the significantly higher amplitude during Interval 8 compared to Interval 3. The enhancement of the propagating disturbance in the plume tends to coincide with its spatial filamentation in both space and time. }
\end{figure}

Examples of signals obtained using the surfing average technique are shown in Figure \ref{fig:waveforms}. The propagating disturbances are substantially stronger in Interval 8 than in Interval 3, which has less spatial filamentation (see \S\ref{sec:structure}). The  amplitude in Interval 8 varies significantly across plume sectors, and is lowest in the middle of the plume ($j=5,6$) where the plumelet edges are the least intense (see Figure\ \ref{fig:roberts_slits} for comparison). 

\begin{figure}[tbh]
\begin{centering}
\includegraphics[width=4.0in]{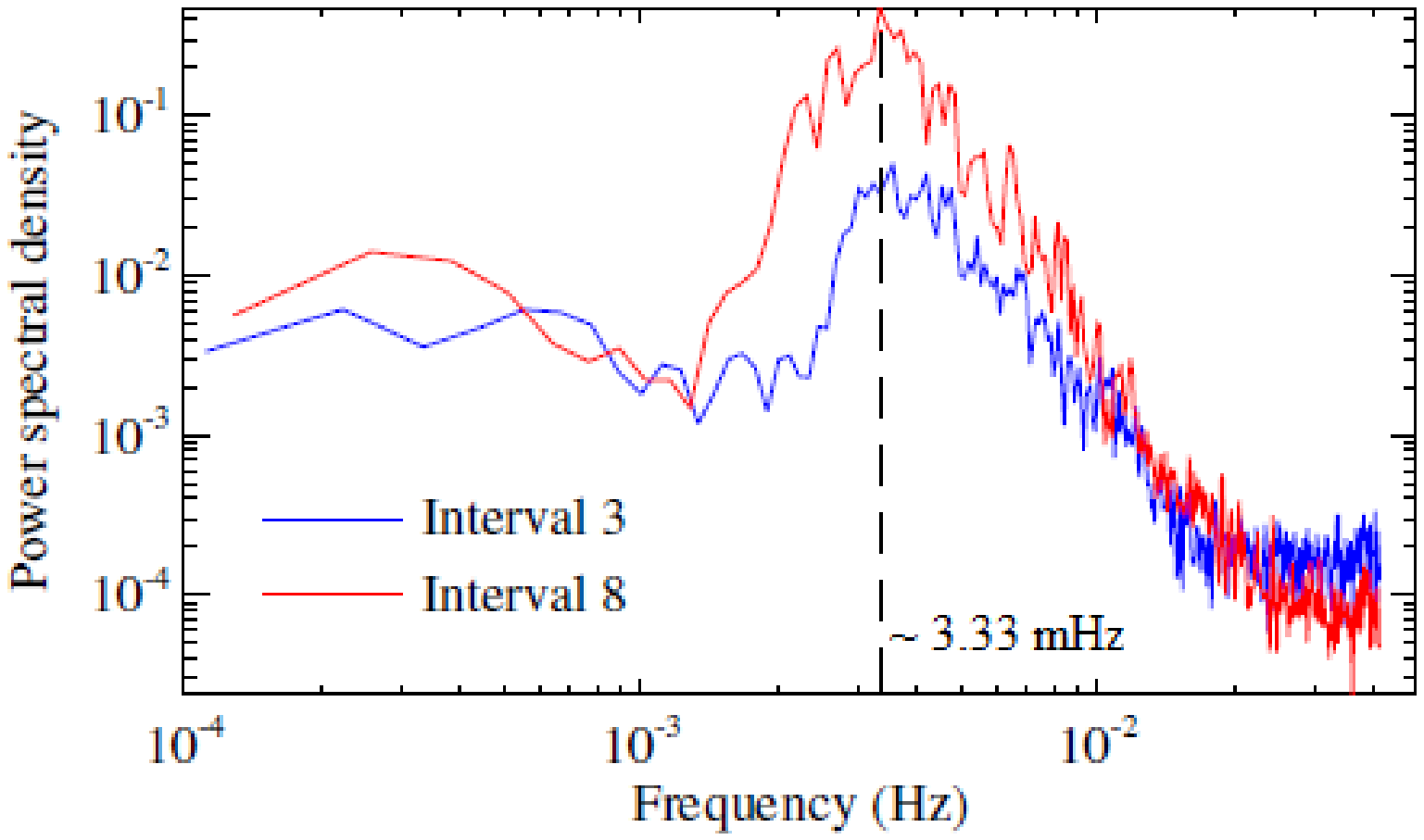}
\par\end{centering}
\caption{\label{fig:spectra} Slit-averaged Fourier power spectra of propagating-front signals during the two intervals shown in Figure \ref{fig:waveforms}. }
\end{figure}

Surprisingly, the low-amplitude oscillations in Interval 3 and the stronger oscillations in Interval 8 have similar shapes and the same well-defined characteristic frequency $\approx$ 3-4 mHz. The Fourier power spectra of the two signals are shown in Figure \ref{fig:spectra}. These spectra were obtained by first Fourier-transforming signals from the individual slits displayed in Figure \ref{fig:waveforms} and then averaging the  power spectra over the slits for a better signal-to-noise ratio. For both intervals, the main spectral peaks are well above the noise floor and have comparable widths. We note that the frequency at the spectral peak is very close to the typical frequency of the $p$-mode oscillations that are ubiquitous in the solar photosphere \citep[see, e.g.,][]{uritsky12}.

We applied the surfing average technique to all nine time intervals of plume observations. Table \ref{table:waves} shows the derived parameters of the dynamic activity. In addition to the apparent velocity $v$, we report the corrected velocity $v_{corr} = v/\cos{(34^\circ)}$, taking into account the average plume angle with respect to the image plane estimated in \S\ref{sec:transverse}. The predicted values of the apparent wavelength $\lambda = vT$ and the corrected wavelength $\lambda_{corr}=v_{corr}T$, corresponding to the measured period $T$, are also provided.

The anomalously long periods in Intervals 1 and 2 reflect an unstable disturbance amplitude during these intervals. Some of the weaker fronts are missed by the method, so the average estimated $T$ value became larger. The other intervals presented in Table \ref{table:waves} are characterized by shorter and mutually consistent $T$ values ranging between $\approx 260$ and $330$ s, with the average period $\left\langle T \right\rangle = 288 \pm 10$ s. The characteristic phase speed varied between $\approx 160$ and $220$ km s$^{-1}$, or over the range $v_{corr} \approx 190$ to $\approx 260$ km s$^{-1}$ after accounting for the projection angle. The average derived wavelength in the image (plume) frame is $50$ ($60$) Mm. The mean values shown in the bottom line of the table exclude Intervals 1 and 2; the reported uncertainties are the standard errors. 

The overall average period $\left\langle T \right\rangle$ reported in Table \ref{table:waves} is in very good agreement with the peak spectral frequency for the examples shown in Figure \ref{fig:spectra}, and is similarly consistent with the dominant $p$-mode oscillation period.

\begin{deluxetable*}{cccccc}
\tablecaption{Parameters of longitudinally propagating disturbances \label{table:waves}}
\tablewidth{0pt}
\tablehead{
\colhead{Interval} & \colhead{$v$ (km s$^{-1}$)} & \colhead{$v_{corr}$ (km s$^{-1}$)} & \colhead{$T$ (s) } & \colhead{ $\lambda$ (Mm) } & $\lambda_{corr}$ (Mm)
}
\decimalcolnumbers
\startdata
0 &      184 $\pm$       11 &      222 $\pm$       13 &      257 $\pm$       11 &       47 $\pm$        3 &       57 $\pm$        4 \\ 
1 &      216 $\pm$       21 &      261 $\pm$       25 &      (734 $\pm$      406) &    $-$ &      $-$ \\ 
2 &      183 $\pm$       18 &      221 $\pm$       22 &     (1416 $\pm$      660) &      $-$ &      $-$  \\ 
3 &      197 $\pm$       17 &      238 $\pm$       20 &      265 $\pm$       21 &       52 $\pm$        6 &       63 $\pm$        7 \\ 
4 &      160 $\pm$       11 &      193 $\pm$       13 &      330 $\pm$       22 &       53 $\pm$        5 &       64 $\pm$        6 \\ 
5 &      164 $\pm$       18 &      198 $\pm$       22 &      281 $\pm$       16 &       46 $\pm$        5 &       55 $\pm$        7 \\ 
6 &      165 $\pm$ 4\phantom{0} &  199 $\pm$ 5\phantom{0} &  326 $\pm$ 9\phantom{0} &   54 $\pm$        2 &       65 $\pm$        2 \\ 
7 &      169 $\pm$       11 &      204 $\pm$       14 &      284 $\pm$ 7\phantom{0} &   48 $\pm$        3 &       58 $\pm$        4 \\ 
8 &      187 $\pm$       19 &      225 $\pm$       23 &      275 $\pm$ 8\phantom{0} &   51 $\pm$        5 &       62 $\pm$        6 \\ 
\hline
$\left\langle \,\,\,\right\rangle$ &     175 $\pm$        5 &      211 $\pm$        6 &      288 $\pm$       10 &       50 $\pm$        1 &       60 $\pm$        1 \\ 
\enddata
\end{deluxetable*}

\subsubsection{Transverse fluctuations}\label{sec:cc}

The concurrent dynamic activity in nearly all plume sectors provides an additional means of evaluating the transverse characteristic scale of the plume substructure, complementary to the methods based on the analysis of static plume images described in \S\ref{sec:structure}. Depending on the underlying physics, the propagating disturbances within morphologically distinct plumelets may or may not be temporally correlated. If the quasi-periodic propagating disturbance is driven by the large-scale dynamics of the photosphere, then one would expect the behavior in different plumelets to be coherent across the plume. If, on the other hand, the activity is driven by the small-scale dynamics occurring above intrusions of minority-polarity magnetic flux sprinkled across the plume base, then one would expect the behavior in different plumelets to be largely independent and not coherent across the plume.

To determine whether collective behavior is at work in the plume, we calculated linear cross-correlation coefficients $C_{jk}$ ($j = 1,...,10$; $k=1,...,10$) characterizing the degree of temporal coherency of propagating signals between all pairs of longitudinal plume slits $j$ and $k$ defined by our adjustable curvilinear coordinate systems (Figure\ \ref{fig:grids}) with zero time lag:
\begin{equation}
    C_{jk} = \frac{\left\langle | S_j S_k | \right\rangle_{t}}{\sigma_j \sigma_k}.
\end{equation}
Here, $S_j$ and $S_k$ are the ``surfing'' signals (\S\ref{sec:dynamics}) in slits $j$ and $k$, obtained using the speed estimates in Table \ref{table:waves}; $\sigma_j$ and $\sigma_k$ are the standard deviations of the two signals; and the angular bracket denotes averaging over the time interval of interest. $C$ is an unsigned measure of temporal coherence that varies between 0 (completely decorrelated behavior) and 1 (perfect correlation). By definition, $C_{jk}=1$ for $j=k$ and the matrix is symmetric ($C_{jk}=C_{kj}$).

\begin{figure}[tbh]
\begin{centering}
\includegraphics[width=5.5in]{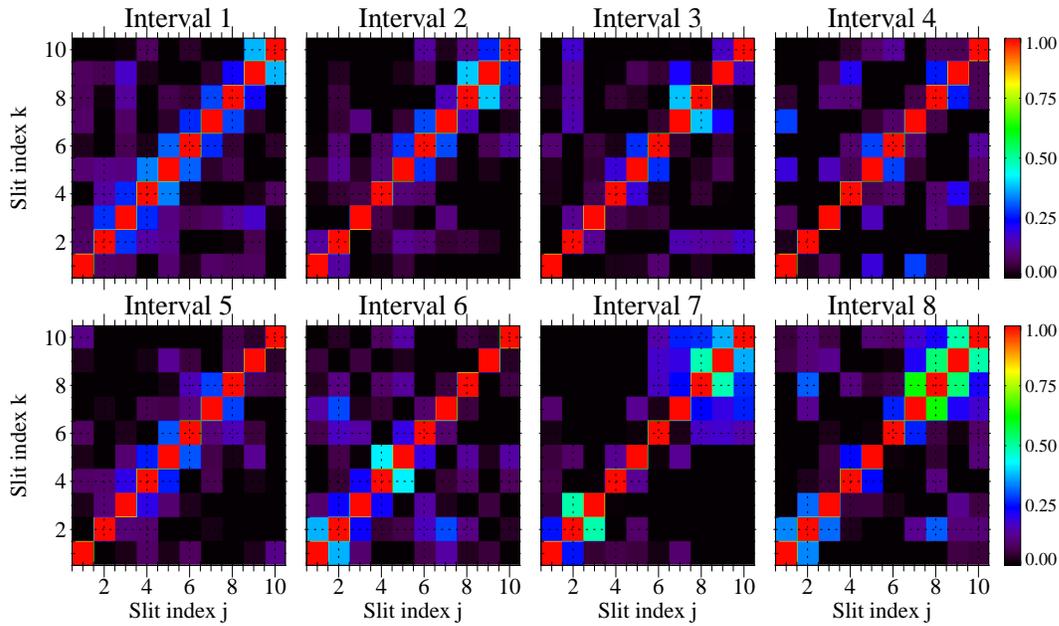}
\par\end{centering}
\caption{\label{fig:cc_arr} Two-dimensional cross-correlation plots showing the linear correlation $C_{jk}$ (Equation \ref{sec:cc}) between the dynamic signals in all pairs of longitudinal slits for the first eight time intervals. Diagonal bins refer to the cross-correlation of each slit with itself, which equals 1 by definition. The correlation coefficient of the adjacent slits tends to be below 0.5 and is effectively 0 for more remote slits, suggesting a characteristic transverse correlation length on the order of slit spacing. The increased correlation scale in slits 7-10 during Intervals 7 and 8 is an artifact caused by the curved shape of those plumelets that crossed the boundaries between the slits near the plume base during this time (see the last panel in Figure\ \ref{fig:grids}).}
\end{figure}

Figure \ref{fig:cc_arr} presents two-dimensional images of the cross-correlation matrices computed for Intervals 1 to 8. The value of $C$ quickly decreases with distance from the main diagonal $j=k$. The two adjacent diagonal lines (sub- and super-diagonals) characterizing the correlation between nearest-neighbor slits ($j=k \pm 1$) typically show $C$ values below 0.5, indicating weak temporal coherence. The slits separated by larger distances show effectively no cross-correlation for most of the time intervals plotted in Figure \ref{fig:cc_arr}. The seemingly longer correlation scale characterizing slits 7-10 during Intervals 7 and 8 is an artifact caused by the curved plumelets crossing the grid boundaries, so that the same wave packet or bulk flow appeared in more than one slit. With this exception, the cross-correlation results suggest that the characteristic length scale describing the interactions between plumelets is not larger than the transverse size of a single longitudinal slit. This size averages 20$''$ and ranges from 8$''$ to 30$''$ depending upon the distance from the plume base and the specific time interval. 

Our scale estimate obviously is affected by the grid size and, therefore, should be considered an upper limit on the actual physical scale controlling interactions between the plumelets. The true correlation length, as suggested by the analysis of the plume morphology (\S\ref{sec:structure}), may be on the order of 10$''$, consistent with the spectral analysis (Figure\ \ref{fig:spatial_spectra}). Alternatively, it may be closer to the measured minimum size of the filamentary substructure, on the order of 1$''$. A definitive determination of the inherent scale of the dynamic behavior and its relationship to plume morphology is an ambitious undertaking that is left for future work.

\section{Discussion}\label{sec:discussion}

We have analyzed a lengthy sequence of observations obtained with the \emph{Solar Dynamics Observatory}'s Atmospheric Imaging Assembly, in which a plume originating in a mid-latitude coronal hole evolved over a few days in July 2016. Nine intervals of virtually continuous, high-resolution images at 12-s cadence were identified; their durations ranged from nearly two to four hours each. We used \emph{noise-gating}  \citep{deforest_noise-gating_2017} and \emph{minsmoothing} \citep{deforest_fading_2016} techniques to process the images, greatly reducing the noise and sharpening the structure to increase the effective resolution of the data set \citep{deforest_structure_2018}. Then we applied the classic edge-detection \citet{roberts1965} transform to identify and track numerous filamentary structures in the resulting \emph{noise-gated, minsmoothed} image sequence. Last, we used a surfing technique \citep{uritsky09,uritsky13,keiling12} to extract durations, periods, and phase speeds from the longitudinal fluctuations of the identified plumelets. Altogether, these new measurements yield key insights into the short- and long-term behaviors of solar plumes.

The most important result of our investigation is that the plume, which appears to be monolithic and more or less uniform at low spatial and temporal resolution, is instead highly structured and dynamic. While earlier studies, including those cited in the Introduction, have reported both substructure and dynamic behavior (waves or flows) in plumes, 
to our knowledge this prior work did not quantitatively assess the longitudinal and transverse structure, oscillation periods and amplitudes, and flow speeds as functions of time in all sectors of the plume, throughout nearly two days of observations. The descriptions to date have largely treated plumes as stationary features subject to perturbations. By more fully analyzing cleaned images of the plume with high spatial and temporal resolution, our investigation reveals that the dynamical portion of the plume is dominant over the stationary structure. This conclusion turns the conventional view of plumes upside down.

We found that the plume brightness was correlated with the number of edges detected (as many as about 30) within the filamentary structure of the plume (see Figure\ \ref{fig:intensity_filamentation}). Over the analysis period, the plumelet count decreased for 15 hours, then increased for 15 hours, while the brightness varied in direct proportion. Thereafter, the count continued to increase for another 10 hours while the brightness remained essentially unchanged. During the very late evolution, the filamentary structure fragmented into more numerous, generally narrower structures (see Intervals 6 through 8 in Figure\ \ref{fig:roberts_transform}). 

In addition to counting the number of contained edges as a quantitative measure of substructural complexity, we measured their characteristic spatial scale (see Figure\ \ref{fig:spatial_spectra}). The average plumelet width is about 10$''$ ($\approx$ 7 Mm). This is commensurate with the average size of minority-polarity intrusions of magnetic flux, EUV and X-ray bright points, and the width of EUV and X-ray jets within coronal holes. These features range in size from 2-3 times our nominal plumelet width down to scales that are much smaller, and generally they are quite numerous. Previous studies have suggested that small-scale coronal-hole jets \citep[e.g.,][]{mcintosh2010} and even smaller-scale, more-transient jetlets \citep[e.g.,][]{raouafi2008, raouafi2014} are major sources of plume mass and kinetic energy. The agreement in the characteristic spatial scales of these features and our plumelets is consistent with the hypothesized connections between plume formation and persistence and the much more dynamic jets and jetlets. To test this relationship in further detail, we are performing a separate analysis of the small-scale, transient brightenings at the base of the observed plume and their associations with the plumelets reported here \citep[][in preparation]{kumar2020}.

By aligning grids along and across the overall plume orientation (Figure\ \ref{fig:grids}), detrending the longitudinal data in time to remove the background slow evolution (Figure\ \ref{fig:four_intervals}), and performing a surfing-average analysis of the residual signals (Figure\ \ref{fig:surfing}), we determined characteristic values of about 1500 s and 200 km s$^{-1}$ for the duration and speed, respectively, of the disturbances (see Table \ref{table:waves}). The former is typical for the duration of larger-scale coronal-hole jets; the latter is typical of the dense outflows in such jets. This agreement with jet lifetimes and speeds strengthens the case for direct links between jets/jetlets and plumes. At the same time, however, the measured speeds of the longitudinal disturbances are fully consistent with their being slow magnetosonic waves. In early analyses of observations from the \emph{Solar and Heliospheric Observatory}'s Extreme-Ultraviolet Imaging Telescope, \citet{ofman97} and \citet{deforest1998} identified similar long-lived, slow-moving disturbances in coronal holes and plumes as such waves. The surfing-average analysis reported here provides additional compelling evidence that time-periodic wave activity occurs on the plumelets (see Figure\ \ref{fig:waveforms}).

The dichotomy of jet-like versus wave-like behavior of the observed plumelets can be resolved straightforwardly if the plumelets are generated by jets and jetlets at the plume base. Each plumelet extends along the plume at a rate determined by the speed of the outflowing jet plasma, and the lifetime of the plumelet structure is determined by the duration of the underlying jet from the source. Clearly, the dense jet is a compressible flow, which would be expected to generate slow magnetosonic waves. The resulting dense plumelet also could support waves generated by external forcing mechanisms. 

Our surfing-average analysis further determined the characteristic oscillation periods of the longitudinal fluctuations in our plumelets. We found an average period across the entire data set of $288 \pm 10$ s (see Table \ref{table:waves}). By taking power spectra of the signals along individual plumelets to extract their frequencies, we found a prominent peak near the corresponding frequency of 3.33 mHz (see Figure\ \ref{fig:spectra}). These results strongly suggest a connection with the pressure-driven $p$-modes that flex the solar photosphere, although in principle this agreement could be coincidental. For the 10-Mm characteristic width of the plumelets, the corresponding latitudinal and longitudinal standing-mode numbers are $\ell,m \approx 200$. Substantial power is present in the $p$-mode spectrum at these and shorter wavelengths 
\citep[e.g.,][]{duvall1997}. Hence, the photospheric undulations driven by the $p$-modes can be much smaller than the plume base under study, and comparable in size to the plumelets and to the typical source regions of coronal-hole jets. The undulation periods ($\approx$ 300 s) are significantly shorter than the typical total duration of the plumelets and the jets ($\approx$ 1500 s). This suggests a scenario in which each plumelet source region rides up and down multiple times on the $p$-mode waves as the jet/jetlet is generated and the plumelet is formed. Consequently, the dense plasma that becomes a distinct plumelet oscillates at the frequency of the underlying photospheric oscillations, and this modulation appears in the plumelet wave signals.

The cross-correlation analysis of different longitudinal strips demonstrates that the plumelet waves were not correlated at transverse scales of 20$''$ (15 Mm) and above (see Figure\ \ref{fig:cc_arr}). This is consistent with the $p$-modes providing the photospheric modulation, whose spatial variations occur on scales as small as the typical separation between the plumelets. Although the underlying frequency of the $p$-modes is common to all plumelets, there is no fixed relationship between the phases of the observed oscillations in any one plumelet and in its neighbor. Hence, the plumelet wave motions are uncorrelated.

High-cadence, high-resolution observations of coronal-hole jets, especially those that exhibit helical outflows along the jet spire, provide strong evidence that nonlinear Alfv\'en waves accompany jet launch and lead the flow of dense plasma into the outer corona \citep[e.g.,][]{kumar2019}, consistent with our jet simulations \citep[e.g.,][]{karpen2017,uritsky17,roberts18}. Our analysis did not reveal any Alfv\'en waves traveling along the plumelets. This is not a strong null result, however, nor is it very surprising, for three reasons. First, as mentioned already, the outflow of dense plasma trails the Alfv\'en waves at the jet front, and increasingly so at higher altitudes due to their speed differential. The dense plasma flow should be expected to provide the main contribution to emission from the elongating plumelet. Second, although an Alfv\'enic shock wave may lead the entire procession from the jet source region, the density enhancement at the shock is not necessarily large, and the Alfv\'en wave itself is incompressible even though it is nonlinear. The contribution of the Alfv\'en wave to the plume emission, therefore, may be insignificant. Third, Alfv\'enic undulations of the plumelets could be difficult to detect in the relatively low-cadence (12-s) \emph{SDO}/AIA observations. For a typical coronal Alfv\'en speed of 1 Mm s$^{-1}$ and a source-region size of 10 Mm or less, the expected Alfv\'en-wave periods are 10 s and below, so these high-frequency waves cannot be properly resolved by \emph{SDO}/AIA.

Our results strongly indicate that the filamentary structure is imprinted dynamically on the plume at its base, and that this structure and dynamics persist out to at least 100 Mm above the surface. It seems unlikely that this coherence can be explained by any mechanism other than its association with the structure of the plume magnetic field, which guides the plasma outflow and the plasma waves. So guided, the plumelet structure and dynamics that we have observed may persist well out into the heliosphere, perhaps to where they can be measured by \emph{Parker Solar Probe} \citep[\emph{PSP};][]{fox_2016} and/or \emph{Solar Orbiter} \citep[\emph{SolO};][]{mueller_2013}. The plumelets' plasma density is enhanced relative to that in the surrounding open corona, and this enhancement should be sustained in the absence of a mechanism that selectively accelerates and rarefies the gas at the front of the plumelet. The characteristic transverse scale ($\sim 10$ Mm) of the plumelets in the corona should expand geometrically, due both to the diverging spherical geometry of space ($\propto R/R_S$) and by the square root of the super-radial factor ($\approx 2$-$4$) by which the area of open magnetic flux on the solar surface expands to fill all 4$\pi$ sr of space sufficiently far from the Sun. The anticipated scale thus ranges from roughly 200 Mm at 10 $R_S$ up to 600 Mm at 30 $R_S$. It is an enticing prospect that \emph{PSP} and/or \emph{SolO} might directly detect the 5-minute oscillations imprinted on the plumelets. If our identification of the solar $p$-modes as the driving mechanism for these waves is correct, this would mean a direct detection of the global dynamics of the solar interior at least 10 $R_S$ out into the heliosphere.  We look forward to seeing the exciting new data that come back from these missions and comparing them to our expectations, which are founded on the detailed analysis of a coronal plume that was observed so diligently by \emph{SDO}.

\section{Conclusions}\label{sec:conclusions}

Filamentary structures and motions in plume images have been reported for many years (e.g., \citet{raouafi2014} and refs therein). Here we presented 
the first in-depth quantitative investigation of these structures, which we denote ``plumelets". Using a large set of high-resolution, high-cadence solar coronal images, we quantified the highly dynamic nature of the plumelets, and demonstrated that their impulsive behavior may, in fact, dominate the large- scale behavior of the ``host'' plume.

We processed almost 40 hours of nearly continuous observations of a typical solar coronal plume by \emph{SDO}/AIA on 2016 July 2-3. Image-processing, edge-detection, and signal-analysis techniques enabled us to obtain the following results:

\begin{enumerate}

\item The plume comprised numerous (10 to 15) filamentary substructures, which we refer to as ``plumelets'', that accounted for most of the variable plume brightness over its lifetime.

\item The width, length, and duration of the plumelets averaged approximately 10 Mm, 100 Mm, and 1500 s, respectively.

\item The plumelets supported persistent longitudinal fluctuations whose speed, period, and wavelength peaked at approximately 200 km s$^{-1}$, 300 s, and 60 Mm, respectively.

\item The longitudinal fluctuations were uncorrelated at transverse scales of 20 Mm and above, i.e., from one plumelet to another.

\item The plumelet oscillation period agrees very closely with the peak-power period of the solar $p$-modes ($\sim$3-5 min).

\item The plumelet width, duration, and speed are consistent with those of the dense outflows in typical coronal-hole jets.

\end{enumerate}

{We are currently investigating the suggestive connections between the plumelets and jetlet activity observed by SDO at the plume base.}

\acknowledgements{The authors are grateful to NASA and the RAS for supporting our research, principally through NASA H-SR and H-ISFM grants to investigate the jet/plume connection and the nature of explosive energy release in the solar atmosphere (VMU, CED, JTK, CRD). Additional support was provided to PK by NASA's NPP program and NASA HGI grant 80NSSC20K0265, to NER by NASA funding for the \emph{PSP} mission, and to PFW by the RAS's fellowship program. We thank the referee for insightful suggestions that have improved the paper.}


\end{document}